\documentclass[11pt,a4paper,oneside,notitlepage]{article}
\pdfoutput=1
\usepackage{jheppub}
\usepackage{hyperref}
\usepackage{graphicx,xcolor}
\usepackage{sectsty,amssymb,amsmath}
\usepackage{enumerate}
\usepackage{booktabs}
\usepackage{multirow}
\usepackage{slashed}
\usepackage[utf8]{inputenc} 
\usepackage{subcaption}
\usepackage{siunitx}
\usepackage{here}
\usepackage{cleveref}
\usepackage{tikz}
\usepackage[normalem]{ulem}

\definecolor{rosso}{cmyk}{0,1,1,0.4}
\definecolor{rossos}{cmyk}{0,1,1,0.55}
\definecolor{rossoc}{cmyk}{0,1,1,0.2}
\definecolor{blu}{cmyk}{1,1,0,0.3}
\definecolor{blus}{cmyk}{1,1,0,0.6}
\definecolor{bluc}{cmyk}{1,1,0,0.1}
\definecolor{verde}{cmyk}{0.92,0,0.59,0.25}
\definecolor{verdec}{cmyk}{0.92,0,0.59,0.15}
\definecolor{verdes}{cmyk}{0.92,0,0.59,0.4}
\definecolor{grigio}{cmyk}{0,0,0,0.07}
\definecolor{rosa}{cmyk}{0,0.1,0.1,0.02}
\definecolor{rosino}{cmyk}{0,0.05,0.05,0.02}
\definecolor{rosas}{cmyk}{0,0.3,0.25,0.05}
\definecolor{celeste}{cmyk}{0.1,0,0,0.02}
\definecolor{giallino}{cmyk}{0,0,0.4,0.02}
\definecolor{rosso}{cmyk}{0,1,1,0.4}
\definecolor{rossos}{cmyk}{0,1,1,0.55}
\definecolor{rossoc}{cmyk}{0,1,1,0.2}
\definecolor{blu}{cmyk}{1,1,0,0.3}
\definecolor{bluc}{cmyk}{1,1,0,0.1}
\definecolor{blucc}{cmyk}{0.7,0.5,0,0}
\definecolor{viola}{cmyk}{0,1,0,0.6}
\definecolor{viola2}{cmyk}{0,1,0.2,0.6}
\definecolor{verde}{cmyk}{0.92,0,0.59,0.25}
\definecolor{verdec}{cmyk}{0.92,0,0.59,0.15}
\definecolor{verdes}{cmyk}{0.92,0,0.59,0.4}
\definecolor{verdino}{cmyk}{0.12,0,0.09,0.05}
\definecolor{giallo}{cmyk}{0,0,1,0}
\definecolor{gialloverde}{cmyk}{0.44,0,0.74,0}
\definecolor{grey}{rgb}{0.6,0.6,0.6}
\definecolor{fuchsia}{rgb}{1,0,1}

\usepackage{chngcntr}
\counterwithin{figure}{section}
\counterwithin{table}{section}
\counterwithin{equation}{section}

\def\({\left(}
\def\){\right)}
\def\be{\begin{equation}}
\def\ee{\end{equation}}
\def\bes{\begin{subequations}}
\def\ees{\end{subequations}}
\def\bea{\begin{eqnarray}}
\def\eea{\end{eqnarray}}
\def\bry{\begin{array}}
\def\ery{\end{array}}
\def\bit{\begin{itemize}}
\def\eit{\end{itemize}}
\def\ben{\begin{enumerate}}
\def\een{\end{enumerate}}

\def\dst{\displaystyle}

\def\f{\frac}

\usepackage{marginnote}

\notoc 

\begin{document}

\title{The Role of Vector Boson Fusion in the Production of Heavy Vector Triplets at the LHC and HL-LHC}

\author[a]{Michael J.~Baker,}    
\author[b]{Timothy Martonhelyi,}   
\author[b]{Andrea Thamm,}     
\author[c]{Riccardo Torre} 

\emailAdd{michael.baker@unimelb.edu.au}   \emailAdd{tmartonhelyi@student.unimelb.edu.au}    
\emailAdd{andrea.thamm@unimelb.edu.au}
\emailAdd{riccardo.torre@ge.infn.it}

\affiliation[a]{ARC Centre of Excellence for Dark Matter Particle Physics, 
 School of Physics, The University of Melbourne, Victoria 3010, Australia}
\affiliation[b]{School of Physics, The University of Melbourne, Victoria 3010, Australia}
\affiliation[c]{INFN, Sezione di Genova, Via Dodecaneso 33, I-16146 Genova, Italy}

\date{\today}

\abstract{We clarify the role of vector boson fusion (VBF) in the production of heavy vector triplets at the LHC and the HL-LHC. We point out that the presence of VBF production leads to an unavoidable rate of Drell-Yan (DY) production and highlight the subtle interplay between the falling parton luminosities and the increasing importance of VBF production as the heavy vector mass increases. We discuss current LHC searches and HL-LHC projections in di-boson and di-lepton final states and demonstrate that VBF production outperforms DY production for resonance masses above $1\,$TeV in certain regions of the parameter space. We define two benchmark parameter points which provide competitive production rates in vector boson fusion.
}

\maketitle
\tableofcontents
\section{Introduction}
\label{sec:introduction}

New heavy vector bosons are present in many well-motivated extensions of the Standard Model (SM). They can appear in weakly coupled models, such as gauge extensions of the SM gauge group, or strongly coupled composite Higgs models. Heavy vectors have been studied extensively in the literature, see for instance refs.~\cite{
Barger:1980ix,
Hewett:1988xc,
Cvetic:1995zs,
Rizzo:2006nw,
Langacker:2008yv,
Salvioni:2009mt,
Salvioni:2009jp,
Chanowitz:2011ew,
Langacker:1989xa,
Sullivan:2002jt,
Grojean:2011vu,
Schmaltz:2010xr,
Frank:2010cj,
Agashe:2007ki,
Agashe:2008jb,
Agashe:2009bb,
Contino:2011np,
Bellazzini:2012tv,
Accomando:2012yg,
CarcamoHernandez:2013ydh,
Low:2015uha,
Accomando:2016mvz,
Liu:2018hum,
Capdevilla:2019zbx,
deBlas:2012tc,
Pappadopulo:2014tg,
Chivukula:2017lyk,
Chivukula:2021foa}, in order to optimise LHC searches and provide the best chances of making a discovery.

The simplified model for heavy vector triplets (HVT) proposed in ref.~\cite{Pappadopulo:2014tg}, together with the accompanying set of computational tools \cite{HVTGitHub}, is useful in motivating, performing, and interpreting searches for heavy vector resonances at the LHC. It provides a framework to set general constraints on colourless heavy vectors that transform as triplets under the SM $SU(2)_{L}$ gauge group (with zero hypercharge) and to interpret the results within a limited parameter space which spans a wide variety of UV-complete models.

In the original reference \cite{Pappadopulo:2014tg} the authors stressed that Drell-Yan (DY) production provides the leading production mechanism for HVT, while vector boson fusion (VBF) is generally suppressed due to the splitting functions multiplying the quark parton distribution functions for weak gauge boson production. Nevertheless, after the energy increase of the LHC from $8$ to $13\,$TeV and with increasing integrated luminosity, the VBF production mode has started to attract the attention of the theoretical \cite{Mohan:2015doa,Florez:2016uob,Cavaliere:2018zcf, Kim:2021vxd}
and experimental communities \cite{
ATLAS:2022jho,
ATLAS:2020fry,
ATLAS:2018sbw,
ATLAS:2018iui,
ATLAS:2018ocj,
ATLAS:2017uhp,
ATLAS:2017otj,
ATLAS:2017jag,
CMS:2022shx,
CMS:2021klu,
CMS:2021itu,
CMS:2021fyk}.
Most of these analyses make the assumption that the HVT can be produced uniquely via VBF. While DY production can be heavily suppressed for very specific values of the parameters, due to a possible cancellation between different contributions to DY production, this assumption is not well-motivated from a UV perspective. In this paper we therefore clarify the role of VBF and show that an ``irreducible'' component of DY production is generically present. We discuss parameter choices that favour VBF production and compare the current and future reach of LHC and HL-LHC searches in DY and VBF production. We demonstrate that the LHC mass reach in VBF becomes competitive for resonance masses around $1\,$TeV and significantly outperforms DY searches at $2\,$TeV.

\section{Heavy Vector Triplet Production: DY and VBF}
\label{sec:production}
We consider a heavy vector triplet, $V^{a}_\mu, \, a= 1,2,3$, transforming as a $(1,3,0)$ under the SM gauge group $SU(3)_c \times SU(2)_L \times U(1)_Y$, in addition to the SM particle content. Using the full Lagrangian in ref.~\cite{deBlas:2012tc} and adopting the notation used in ref.~\cite{Pappadopulo:2014tg}, the relevant Lagrangian terms for this work are
\begin{align}
\dst{\mathcal{L}}_V \supset  & \dst-\frac14 D_{[\mu}V_{\nu ]}^a D^{[\mu}V^{\nu ]\;a}+\frac{m_V^{2}}2V_\mu^a V^{\mu\;a} +\, i\,g_V  c_H V_\mu^a H^\dagger \tau^a {\overset{{}_{\leftrightarrow}}{D}}^\mu H \vspace{2mm}\\
&\dst+\frac{g^2}{g_V} c_q V_\mu^a \sum_q\overline{q_L}\gamma^\mu\tau^a q_L +\frac{g^2}{g_V} c_\ell V_\mu^a \sum_{e,\mu,\tau}\overline{\ell_L}\gamma^\mu\tau^a \ell_L \,,
\end{align}
where 
\begin{equation}
D_{[\mu}V^a_{\nu ]}=D_{\mu}V^a_{\nu } -D_{\nu}V^a_{\mu }\,,\;\;\;\;\; D_\mu V_\nu^a = \partial_\mu V_\nu^a
+g\,\epsilon^{abc}W_\mu^bV_\nu^c\,,
\end{equation}
the Higgs current is given by $i\,H^\dagger \tau^a {\overset{{}_{\leftrightarrow}}{D}}^\mu H=i\,H^\dagger \tau^a D^\mu H\,-\,i\,D^\mu H^\dagger \tau^a  H$, and $\tau^a=\sigma^a/2$. The coupling $g_V$ describes the typical strength of the heavy vector interactions and the $c_i$ parameters denote the deviation from this typical strength. The parameter $c_H$ controls the mixing of the heavy vectors with SM gauge bosons on electroweak symmetry breaking, and their decay widths into SM gauge bosons. The coupling of the HVT to quarks and leptons is mostly controlled by $c_q$ and $c_\ell$, respectively. Note that $g_{V}$ is a redundant coupling since the HVT couplings to fermions are proportional to the combinations $c_{q}/g_{V}$ and $c_{\ell}/g_{V}$, while the coupling to gauge bosons enters as $g_{V} c_{H}$. In the remainder of the paper, we will thus show the parameter space in terms of these coupling combinations. We will also take $c_\ell = c_{\ell 1} = c_{\ell 2} = c_{\ell 3}$ and $c_q = c_{q1} = c_{q2} \neq c_{q3}$ since the HVT coupling to third generation quarks may differ substantially from the couplings to the first two generations. While $g_V c_H$ and $c_{q}/g_V$ will play important roles for both production and decay of the heavy resonances at the LHC, $c_\ell/g_V$ and $c_{q3}/g_V$ will only significantly impact the decay. For further details on the basic phenomenology of this simplified model including physical masses and mixings we refer the reader to ref.~\cite{Pappadopulo:2014tg}.

We now discuss relevant ingredients that enter a comparison between HVT production via DY and VBF. 
We will restrict our attention to cases that can be described by the Narrow Width Approximation (NWA), i.e.~when the total cross-section can be factorised into the production cross-section, $\sigma$, and the decay Branching Ratio (BR). In practice, this is a good approximation for widths less than around 15\% of the particle's mass.  In the case of a large width, a different experimental analysis would be needed and an interpretation in terms of an Effective Field Theory (EFT) would be more suitable. EFT searches in the spectrum of di-leptons, di-jets and di-bosons are the perfect complement to the search for narrow resonances, and explicit new physics models, taking into account the full particle spectrum, can then be used to study any hint of new physics arising from either approach. The production cross-section of a narrow resonance can be written in terms of the partial widths $\Gamma_{V\,\to\, i j}$ of the decay processes $V\to ij$ as
\begin{equation}
\sigma(p p \to V+X) = \sum_{i,j \,\in\, p}\f{\Gamma_{V\,\to \,i j}}{M_{V}} \f{16\pi^{2}(2J+1)}{(2S_{i}+1)(2S_{j}+1)}\f{C}{C_{i}C_{j}}\f{dL_{i j}}{d\hat{s}}\Bigg|_{\hat{s}=M_{V}^{2}}\,.
\label{CS}
\end{equation}
In this equation, $i,j=\{q,\overline{q},W,Z\}$ denote the colliding partons in the two protons and $dL_{ij}/d\hat{s}|_{\hat{s}=M_{V}^{2}}$ describes the corresponding parton luminosities evaluated at the resonance mass. Here we include the $W$ and $Z$ bosons as partons and obtain the relevant parton luminosities by convolving the appropriate quark parton distribution functions with the relevant splitting functions, see ref.~\cite{Pappadopulo:2014tg} for details. The factor $J$ is the spin of the resonance and $C$ its colour factor. $S_{i,j}$ and $C_{i,j}$ are the analogous quantities for the initial states. 

There are three main contributions to this formula. Firstly, the numerical factor
\begin{equation}
\label{CS-2}
    N=\f{16\pi^{2}(2J+1)}{(2S_{i}+1)(2S_{j}+1)}\f{C}{C_{i}C_{j}}\,,
\end{equation}
which only depends on the quantum numbers of the particles involved. Numerically, we find
\begin{equation}
\label{CS-3}
    \displaystyle N_{\text{DY}}=\f{4\pi^{2}}{3}\,, \hspace{1cm}
    \displaystyle N_{\text{VBF}}=48\pi^{2}\,,
\end{equation}
for DY and VBF where we have taken into account the fact that the HVT couples mostly to longitudinal SM gauge bosons and have therefore set $S_{i}=S_{j}=0$. This factor therefore favours VBF with a ratio of $N_{\text{VBF}}/N_{\text{DY}}=36$.

\begin{figure}
    \centering
    \rotatebox{90}{\phantom{XXXXXX}
    \scalebox{0.8}{$\frac{(dL/d\hat{s})_\text{VBF}}{(dL/d\hat{s})_\text{DY}}$}}
    \includegraphics[height=0.32\textwidth]{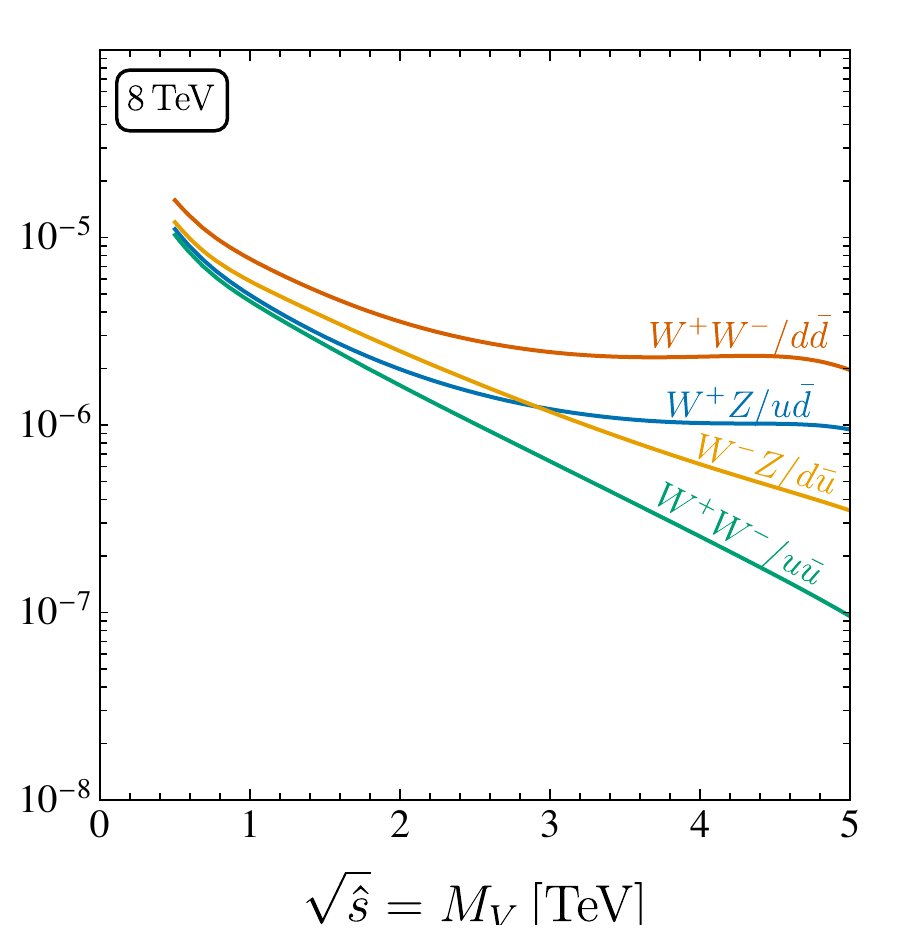}
    \includegraphics[height=0.32\textwidth]{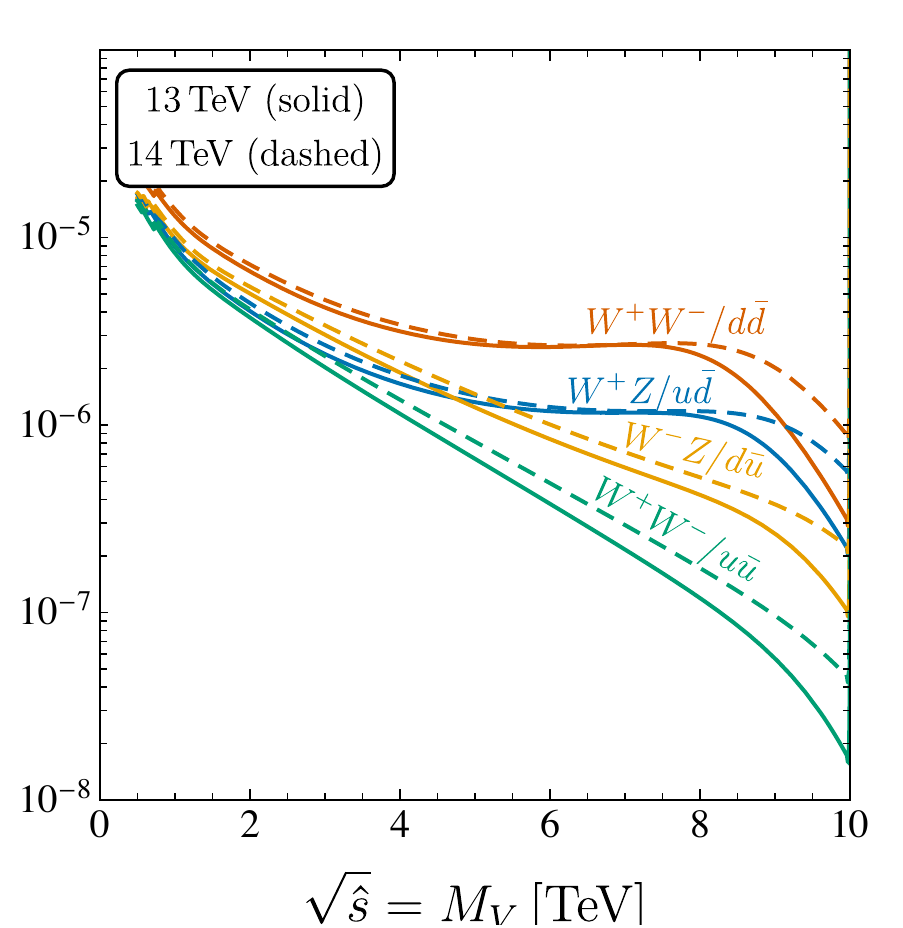}
    \includegraphics[height=0.32\textwidth]{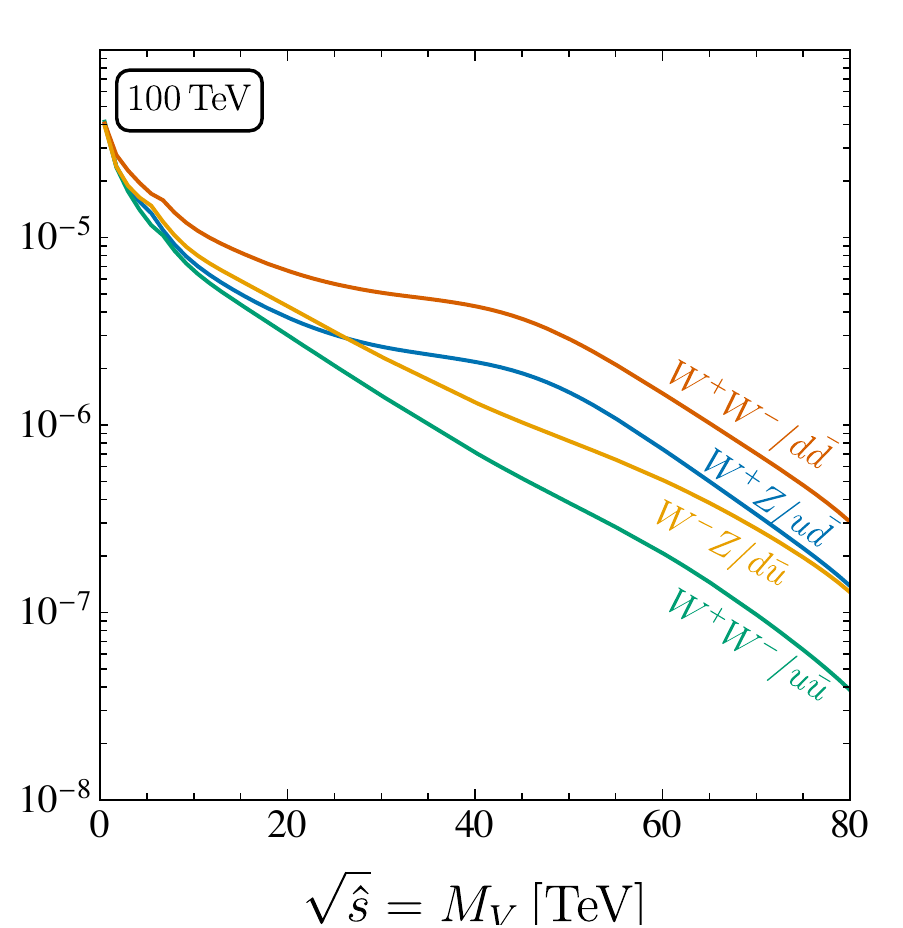}
    \caption{
        The ratio of VBF to DY parton luminosities as a function of the resonance mass for different collider energies. 
        The $W^+W^-$, $d\bar{d}$ and $u\bar{u}$ channels are relevant for $V^0$ production while $W^+Z$ and $u\bar{d}$ ($W^-Z$ and $d\bar{u}$) are relevant for $V^+$ ($V^-$) production.
    }
    \label{fig:pdf-ratio}
\end{figure}

Secondly, the parton luminosities. These are the main source of suppression of resonance production via VBF with respect to DY due to the $\alpha_{\text{EW}}$ suppression entering through the splitting functions. Their values at different collider energies were shown in fig.~2.2 of ref.~\cite{Pappadopulo:2014tg}. In \cref{fig:pdf-ratio} we show the ratio of VBF to DY parton luminosities as a function of the resonance mass for different collider energies. We see that the parton luminosities for VBF are suppressed with respect to DY production by a factor of $10^{-5}$ to $10^{-7}$, depending on the resonance production channel, its mass and the collider energy. This effect significantly outweighs the VBF enhancement due to the numerical factor discussed above.

Whether VBF can become competitive is then down to the third ingredient of \cref{CS}, the decay widths. For VBF to compete with or even overcome DY production, the decay width into two bosons needs to be much larger than the decay widths into two light quarks (while still satisfying the constraint of a narrow resonance). The relevant partial widths are approximately
\begin{align}
\label{eq:widths-1}
\Gamma_{V^\pm\to W^\pm_LZ_L } \simeq &\, \Gamma_{V^0\to W^+_LW^-_L} \simeq \Gamma_{V^\pm\to W^\pm_L h} \simeq \Gamma_{V^0\to Z_L h} \simeq \frac{ g_{V}^{2} c_{H}^{2} M_{V}}{192\pi}\,,
\\
\Gamma_{V^\pm\to q\overline{q}'} \simeq &\, 2\Gamma_{V^0\to q\overline{q}}
\simeq \frac{g^{2}M_{V}}{16\pi}\frac{g^{2}}{g_{V}^{2}}\left[c_{q}^{2}\left(1-c_{H}^2\frac{g^{2}}{g_{V}^{2}}\zeta^{4}\right)+2c_{q}c_{H}\zeta^{2}\left(1+\frac{g^{2}}{g_{V}^{2}}\zeta^{2}\right)+c_{H}^2 \zeta^{4}\right]\,,
\label{eq:widths-2}
\end{align}
where in the first line we have retained terms up to $\mathcal{O}(\zeta)$, where
\begin{equation}
    \zeta\simeq \frac{g_{V}m_{W}}{g M_{V}}\,,
\end{equation}
as in ref.~\cite{Pappadopulo:2014tg}, and in the second line we have taken the limit $g' \to 0$ and retained terms up to $\mathcal{O}(\zeta^4)$.

For $c_q \sim 1$, the decay widths into bosons are not much larger than the decay widths into light quarks (again restricting ourselves to parameters where the narrow width approximation is applicable). The ratio of di-boson to light quark decay widths is
\begin{equation}
\label{eq:widthsratio1}
\frac{\Gamma_{V^\pm\to W^\pm_LZ_L}}{\Gamma_{V^\pm\to q\overline{q}'}}
\simeq \frac{1}{2} \frac{\Gamma_{V^0\to W^+_LW^-_L}}{\Gamma_{V^0\to q\overline{q}}} 
=\frac{1}{12}\frac{g_V^{4}}{g^4} \frac{c_H^2}{c_q^2} + \mathcal{O}(\zeta^2) \,,
\end{equation}
which is of order one for $g_V \sim c_H \sim c_q \sim 1$. As such, VBF production is then suppressed by the parton luminosities compared to DY production.
In ref.~\cite{Pappadopulo:2014tg} it was shown that $c_H/c_q \sim 3$ and $g_V=6$ (or, more generally, $g_{V}^{2}c_{H}/c_{q}\sim 108$) is needed for VBF to overcome DY production at $14\,$TeV. This can only be satisfied, while maintaining a total width less than 15\%, if $c_q/g_V \lesssim 0.05$. Small values of $c_q/g_V$ appear naturally in certain UV models such as walking technicolour~\cite{Belyaev:2018jse,Belyaev:2019ybr}.

We therefore focus on the region of parameter space where $c_q/g_V \ll 1$. For vanishing $c_q=0$, we see from \cref{eq:widths-2} that we retain a non-zero decay width into light quarks. This is due to the HVT coupling to light quarks inherited from the SM gauge bosons through the mixing induced by $c_{H}$. However, the decay width into quarks is now suppressed and the ratio of the widths into gauge bosons and light quarks is given by
\begin{equation}\label{eq:widthsratio}
\frac{\Gamma_{V^\pm\to W^\pm_LZ_L}}{\Gamma_{V^\pm\to q\overline{q}'}^{(c_{q}=0)}}
\simeq \frac{1}{2} \frac{\Gamma_{V^0\to W^+_LW^-_L}}{\Gamma_{V^0\to q\overline{q}}^{(c_{q}=0)}}
\simeq \frac{1}{12}\frac{g_V^{4}}{g^{4}}\zeta^{-4}=\frac{1}{12}\frac{M_{V}^{4}}{ m_{W}^{4}} \,.
\end{equation}
This implies that, in the limit of vanishing $c_q$, the ratio of decay widths, and hence also the ratio of VBF over DY production rates, increases for increasing HVT resonance masses and is independent of any other HVT parameter (up to small corrections). It turns out that this increase grows faster with mass than the reduction in partial width ratios seen in \cref{fig:pdf-ratio}, so VBF production becomes increasingly important for larger resonance masses.

Putting these three ingredients together for example resonance masses of $M_{V}=1 \, (2)\,$TeV at the $13\,$TeV LHC gives us a numerical factor ratio of $36$ as shown in \cref{CS-3}, a ratio of parton luminosities of about $5\cdot10^{-6}$ $(3\cdot10^{-6})$ and a partial widths ratio of $2\cdot 10^{3}$ $(3\cdot 10^{4})$ as given by \cref{eq:widthsratio}. For $c_q=0$, this leads to a ratio of VBF over DY production of $0.4 \, (3.2)$.

\begin{figure}
    \centering
    \rotatebox{90}{\phantom{XXXXXXXXl}
    \scalebox{0.65}{$c_q/g_V$}}
    \includegraphics[height=0.32\textwidth]{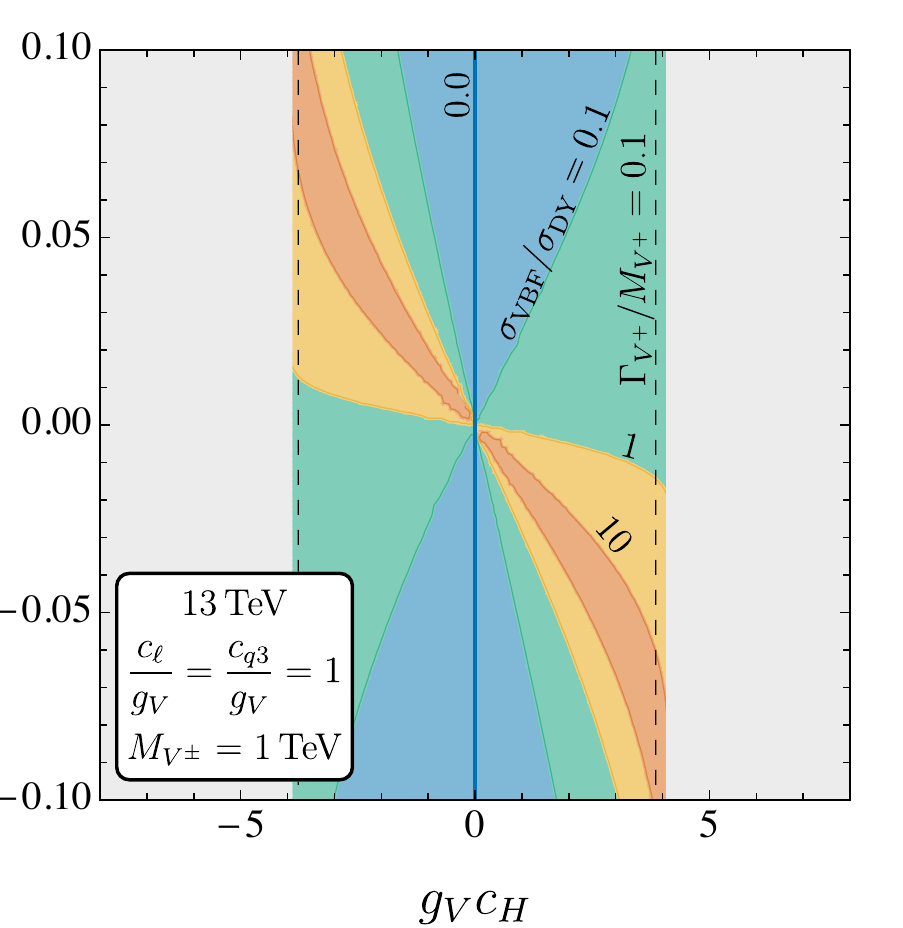}
    \includegraphics[height=0.32\textwidth]{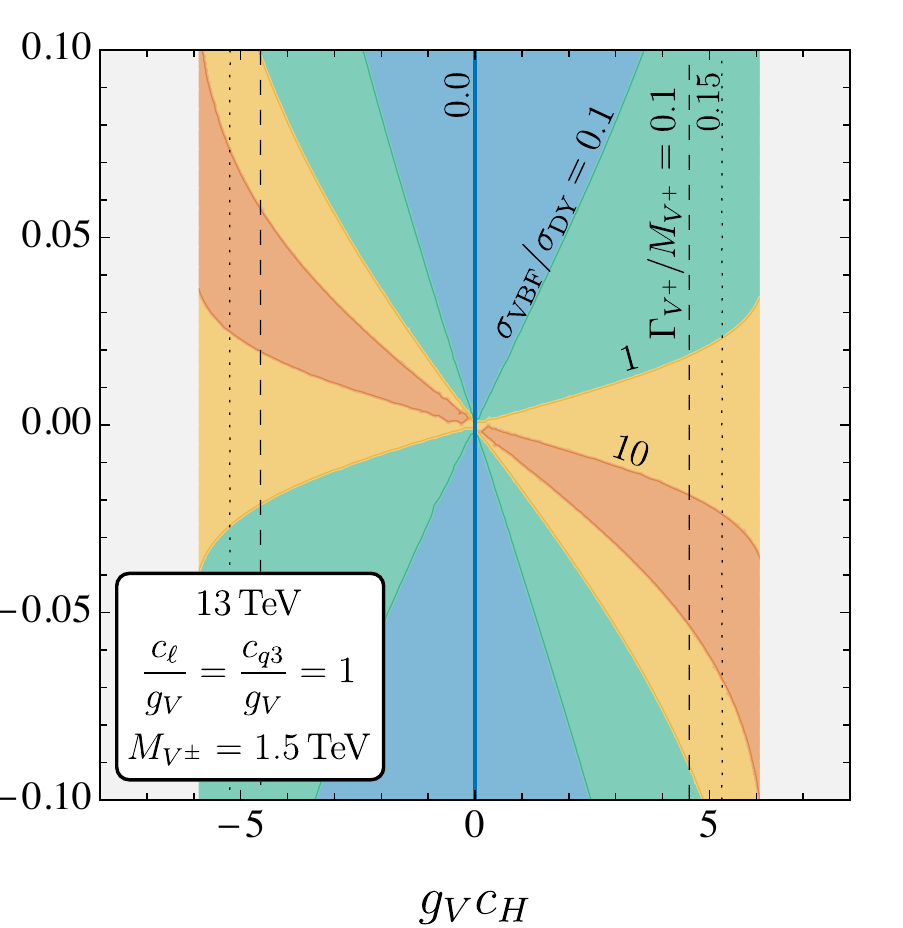}
    \includegraphics[height=0.32\textwidth]{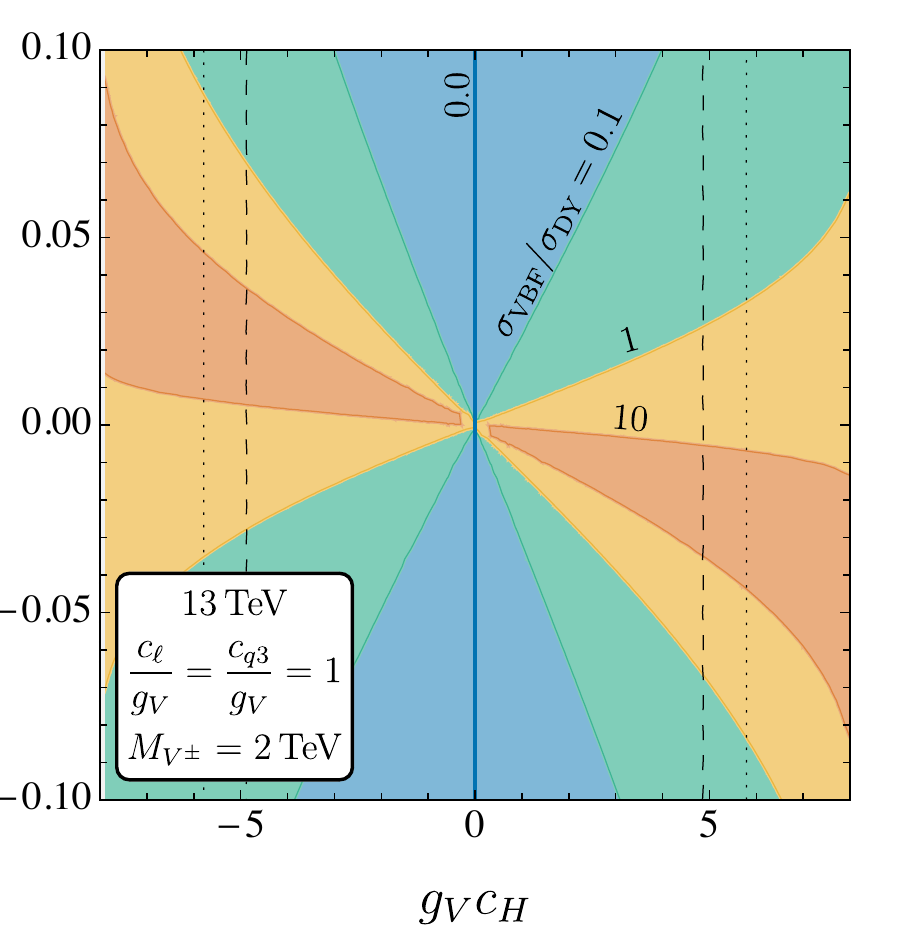}
    \\
    \caption{
        The ratio of VBF to DY production cross-sections for charged resonances in the $g_V c_H-c_q/g_V$ parameter region for resonance masses of $M_{V^\pm} = 1$, $1.5$, and $2\,$TeV (left, centre and right) at $13\,$TeV.  In the blue and red regions the ratios become $\ll 1$ and $\gg 1$, respectively. Theoretically disallowed regions in the simplified model are shown in light grey.
    }
    \label{fig:ratio}
\end{figure}

This behaviour is further exemplified in \cref{fig:ratio}, which shows the ratio of VBF to DY production cross-sections in the $g_V c_H - c_q/g_V$ parameter space for charged resonance masses of $M_{V} = 1, 1.5$ and $2\,$TeV. We see that the parameter space where VBF outperforms DY production (the red and yellow regions, where $\sigma_\text{VBF}/\sigma_\text{DY} > 1$) is confined to the region $|c_q|/g_V \lesssim 0.1$ but that its area increases for increasing resonance mass. The largest $\sigma_\text{VBF}/\sigma_\text{DY}$ ratios are obtained when DY production is smallest. As can be seen from \cref{eq:widths-2}, in the limit $g'=0$ and at $\mathcal{O}(\zeta^2)$, $\Gamma_{V^\pm\to q\overline{q}'}$, and hence the DY production cross-section, vanishes for 
\begin{equation}
\label{eq:cq-ch-relation}
    c_q = -c_H \zeta^2 + \mathcal{O}(\zeta^3) \,.
\end{equation}
The $\sigma_\text{VBF}/\sigma_\text{DY}$ ratio can be enhanced if $c_H$ and $c_q$ have opposite signs and the appropriate magnitude. 
However, there is generally no reason that \cref{eq:cq-ch-relation} should be satisfied in a complete UV theory, since the parameters $c_{q}$ and $c_{H}$ have completely different UV origins. Instead, to be conservative, we will take $c_q = 0$ in the remainder of this work.
The light grey areas in \cref{fig:ratio} depict theoretically forbidden regions, where the values of the SM parameters cannot be reproduced in the HVT simplified model (although these regions of parameter space could potentially become viable in UV complete theories). In our model, real Lagrangian parameters can only lead to real observable quantities, such as masses, if $|g_V c_H| < g^2(m_{V^0}^2 - m_Z^2)/2 m_Z m_{V^0} \sqrt{g^2-4\pi \alpha} $.\footnote{We will see later that the condition has a weak dependence on $c_\ell$, which enters as $g$ depends on $c_\ell$.}  In \cref{fig:ratio} and in all figures in this work, the theoretically allowed regions satisfy electroweak precision tests at $2\sigma$ for HVT masses above $1\,$TeV. We see that as the resonance mass increases, larger values of $g_V c_H$ become viable (although care should always be taken to ensure that perturbative unitarity is not violated). The dashed and dotted black lines show where the total width of the resonance is 10\% and 15\%, respectively.

\begin{figure}
    \centering
    \includegraphics[width=0.5\textwidth]{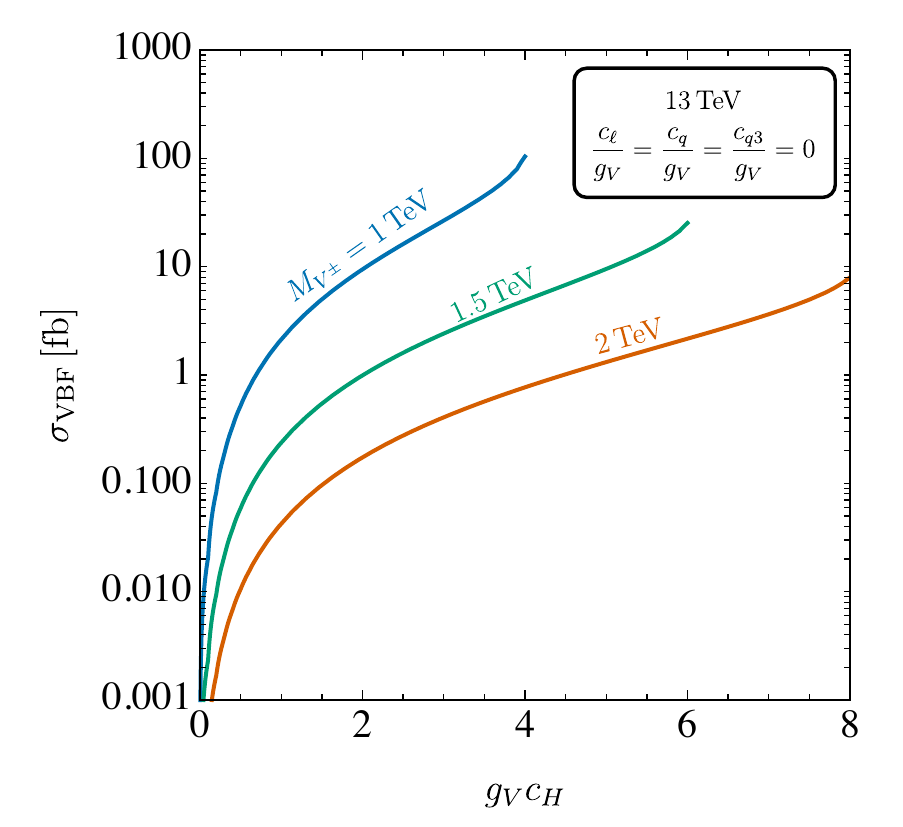}
    \\
    \caption{The VBF production cross-section of $V^++V^-$ as a function of $g_V c_H$ for resonance masses of 1, 1.5 and 2 TeV.
    }
    \label{fig:sigma-ch}
\end{figure}

As the HVT mass grows, the corresponding parton luminosities decrease rapidly. At larger masses the production cross-section reduces and VBF searches will eventually lose sensitivity. This can also be seen in \cref{fig:sigma-ch} which shows the VBF production cross-section vs $g_V c_H$ for masses of $1, 1.5$, and $2\,$TeV. Note that $g_V c_H \ll 1$ is also very difficult to probe, again because the production cross-section is small.

Finally, note that $c_{\ell}$ and the coupling to third generation quarks, $c_{q3}$, have a small but non-zero impact on the VBF over DY cross-section ratio. This is due to $c_{\ell}$ entering the definition of the Fermi decay constant, $G_{F}$, which we use as an input parameter, extracted from muon decay, and to both $c_\ell$ and $c_{q3}$ entering the total width. Nevertheless, the impact of these two couplings on the production cross-section ratio is small (up to about $10\%$) and can generally be neglected for our purposes.

\section{Heavy Vector Triplet Decay}
\label{sec:decay}

Once produced, the HVT can decay into light or heavy quarks, di-leptons or di-bosons. In different regions of parameter space the relative importance of each decay channel varies dramatically. As discussed in the previous section, here we will focus on the region where $c_q/g_V \ll 1$. We argued that the presence of the mixing parameter $c_H$ leads to partial widths into di-bosons proportional to $g_V^2 c_H^2$ and a partial width into quarks proportional to $g_V^2 c_H^2 m_W^4/M_V^4$. This parametric behaviour, along with $m_W \ll M_V$, implies that di-boson decays always dominate over di-jet final states. Decay into di-jets is therefore subdominant and generally irrelevant for VBF studies.

\begin{figure}
    \centering
    \includegraphics[width=0.49\textwidth]{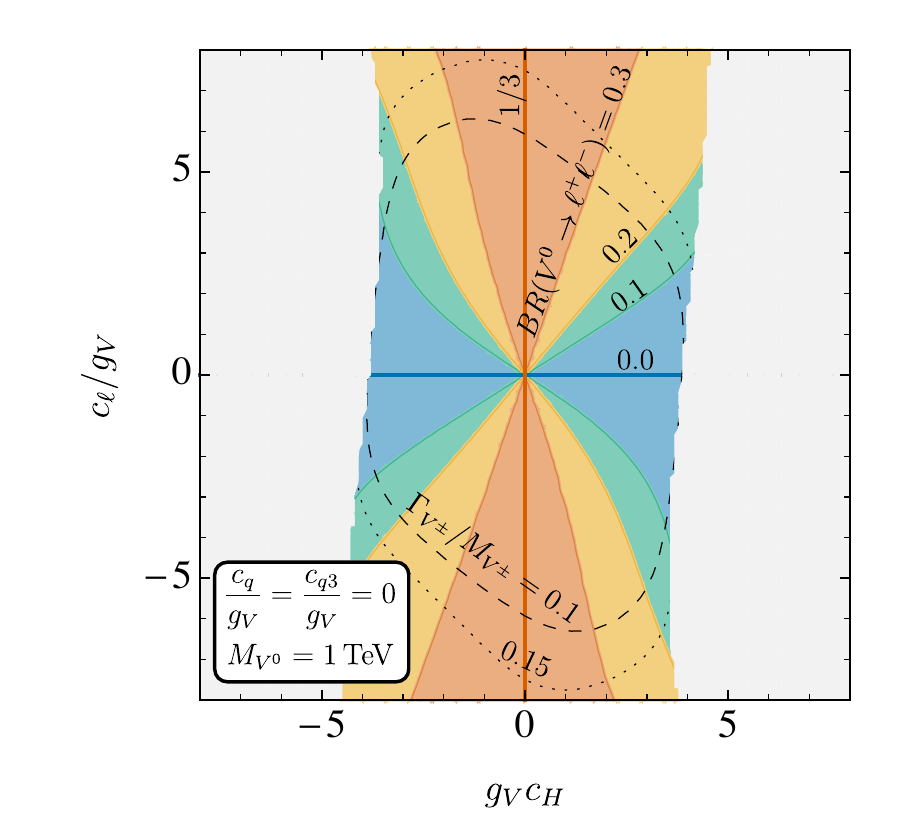}
    \includegraphics[width=0.49\textwidth]{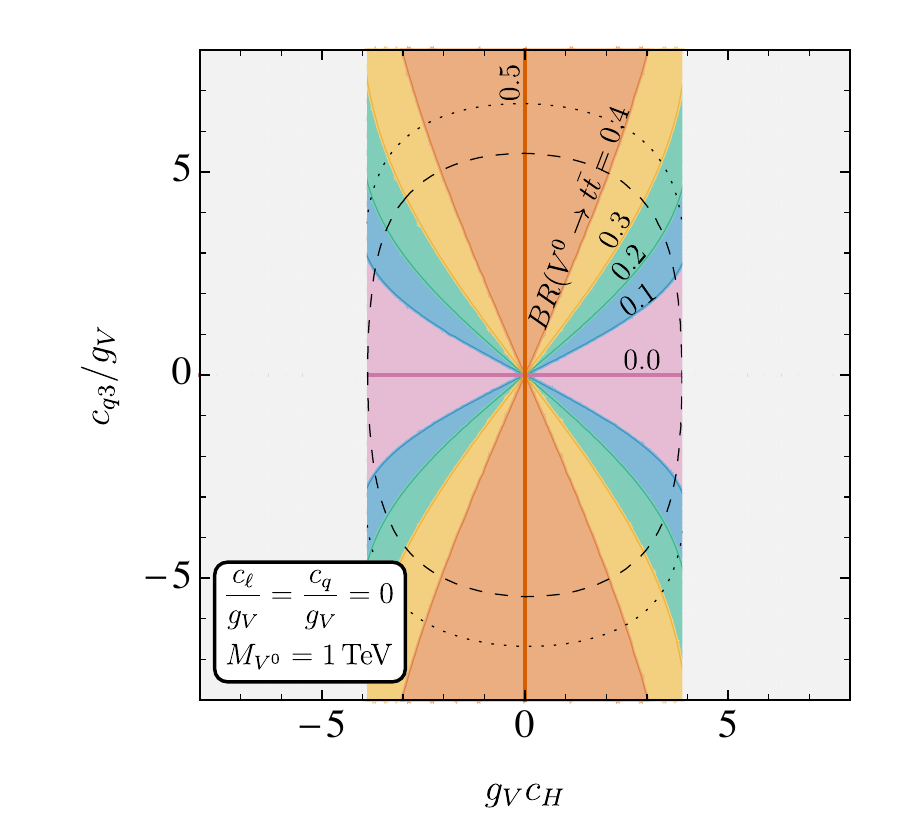}
    \caption{
        The branching ratios of $V^0$ into $\ell^+\ell^-$ for $\ell = e$ or $\mu$ (left) and into $t\bar{t}$ (right) for $M_{V^0}=1\,\text{TeV}$. The dashed and dotted black lines depict contours of constant total width corresponding to $10\%$ and $15\%$ of the resonance mass.
    }
    \label{fig:br-vz}
\end{figure}

Di-lepton decays enter with an independent coupling, $c_\ell$, and the corresponding widths are given by
\begin{equation}\label{eq:widthsdilep}
\dst \Gamma_{V^\pm\to \ell\overline{\ell}'} \simeq 2\Gamma_{V^0\to \ell\overline{\ell}}
\simeq \left(\frac{g^2 c_{\ell}}{g_V}\right)^2 \frac{M_{V}}{48\pi}\,.
\end{equation}
The decay into di-leptons can be comparable to or dominate over di-boson decays depending on the relative size of $c_\ell$ and $c_H$. \Cref{fig:br-vz} (left) shows the branching ratios of the neutral component of the HVT into $e^+e^-$ or $\mu^+\mu^-$ for a resonance mass of $1\,$TeV. As expected, the branching ratio into di-leptons is largest for sizeable values of $c_\ell/g_V$ and small $g_V c_H$. Note that $c_\ell$ controls the decays into charged leptons and neutrinos. When $c_H=c_q=c_{q3}=0$ the branching ratio into leptons is 1, and the branching ratio into $e^+e^-+\mu^+\mu^-$ is 1/3. The branching ratio into di-bosons is just one minus this plot (when all flavours and the neutrinos are taken into account) and dominates for small values of $c_\ell/g_V$ and large $g_V c_H$. 

The coupling to third generation quarks, $c_{q3}$, can in general be different to the coupling to light quarks. In \cref{fig:br-vz} (right) we show the branching ratio into $t\bar t$, assuming $c_\ell = c_q = 0$ (which will be equal to the branching ratio to $b\bar{b}$, up to small corrections due to the available phase space). As in the di-lepton case, this branching ratio is largest for large $c_{q3}/g_V$ and small $g_V c_H$, while di-boson decays dominate at small $c_{q3}/g_V$ and large $g_V c_H$.  However, since searches for third generation quarks are typically less constraining than di-lepton searches, in what follows we will focus on non-zero $c_l$ and the associated di-lepton signatures rather than non-zero $c_{q3}$.

The dashed and dotted black lines in \cref{fig:br-vz} again depict contours of constant total width, at 10\% and 15\% of the resonance mass. Demanding a narrow resonance constrains $|c_\ell|$ $(|c_{q3}|)$ to values below $6$ ($5$) or $8$ $(6)$ for $\Gamma_\text{tot}/M_V = 0.1$ and $0.15$, respectively. The mixing parameter is constrained to be $|c_H| \lesssim 4$ for a resonance mass of $1\,$TeV, otherwise the SM parameters cannot be reproduced in the HVT simplified model.

\section{Current and Future Limits}
\label{sec:limits}

Before examining the impact of current limits on the $g_V c_H$--$c_\ell/g_V$ parameter space, we propose two benchmark parameter points of the HVT model which favour resonance production via VBF:

\begin{itemize}
\item {\bf VBF-DB (Di-Boson) Benchmark:} $g_V c_H = 4$, $c_\ell/g_V =0$, $c_q/g_V = c_{q3}/g_V = 0$ \\
These benchmark parameters provide dominant resonance production via VBF for masses $\gtrsim 1\,\text{TeV}$ and almost total decay into di-bosons. Note that this benchmark is at the threshold of both theoretical consistency and current indirect constraints from electroweak precision tests for $m_V \sim 1\,\text{TeV}$, although both of these constraints relax at higher masses and may be modified in complete UV models.

\item {\bf VBF-DL (Di-Lepton) Benchmark:} $g_V c_H = 3$, $c_\ell/g_V =-3$, $c_q/g_V = c_{q3}/g_V = 0$ \\
These benchmark parameters provide competitive resonance production via VBF and reasonable decay into di-leptons. While branching ratios into di-bosons are still sizeable, current di-boson searches are less sensitive than di-lepton searches at this benchmark point. Decays into quark final states are negligible. We choose opposite signs for $g_V c_H$ and $c_\ell/g_V$ since this slightly improves the VBF to DY production ratio, due to sub-leading effects discussed at the end of \cref{sec:production}. This benchmark is only theoretically consistent for $M_V \gtrsim 0.8\,\text{TeV}$.
\end{itemize}

Recent experimental analyses \cite{ATLAS:2022jho,CMS:2021fyk, ATLAS:2020fry,CMS:2021klu,CMS:2021itu} defined the HVT parameters $g_V=1$, $c_H = 1$, $c_\ell =0$, $c_q = c_{q3} = 0$, sometimes referred to as ``Model C'', as a benchmark for their VBF analyses. While this choice of parameters is very similar to our VBF di-boson benchmark, our choice of $g_V c_H = 4$ increases the VBF production cross-section while retaining the validity of the narrow width approximation. We furthermore want to stress that $c_q = 0$ does not imply vanishing DY production. As discussed in \cref{sec:production}, DY production will also be induced via mixing through non-zero values of $c_H$.

\subsection{Current LHC Limits}
\label{subsec:LHClimits}

 \begin{table}[t]
    \begin{center}
    {
    \begin{tabular}{c|c}
    \hline
    \hline
    Channel & Reference \\
    \hline 
    $WZ \to \ell \nu \ell' \ell'$ & \cite{ATLAS:2022jho}\\
    $Zh \to \text{leptons}~\text{hadrons}$ & \cite{CMS:2021fyk}\\
    $WW,WZ \to \text{leptons}~\text{hadrons}$ & \cite{ATLAS:2020fry,CMS:2021klu,CMS:2021itu}\\
    \hline
    $\ell\ell$ & \cite{ATLAS:2019erb,CMS:2021ctt}\\
    $\ell\nu$ & \cite{ATLAS:2019lsy,CMS:2022yjm}\\
    $\tau\nu$ & \cite{ATLAS:2021bjk}\\
    \hline 
    \hline
    \end{tabular}
    }
    \caption{ATLAS and CMS di-boson and di-lepton searches considering DY and VBF production with a luminosity $\sim 140\,\text{fb}^{-1}$.}
    \label{tab:expsearches}
    \end{center}
\end{table}

\begin{figure}
    \centering
    \includegraphics[width=0.49\textwidth]{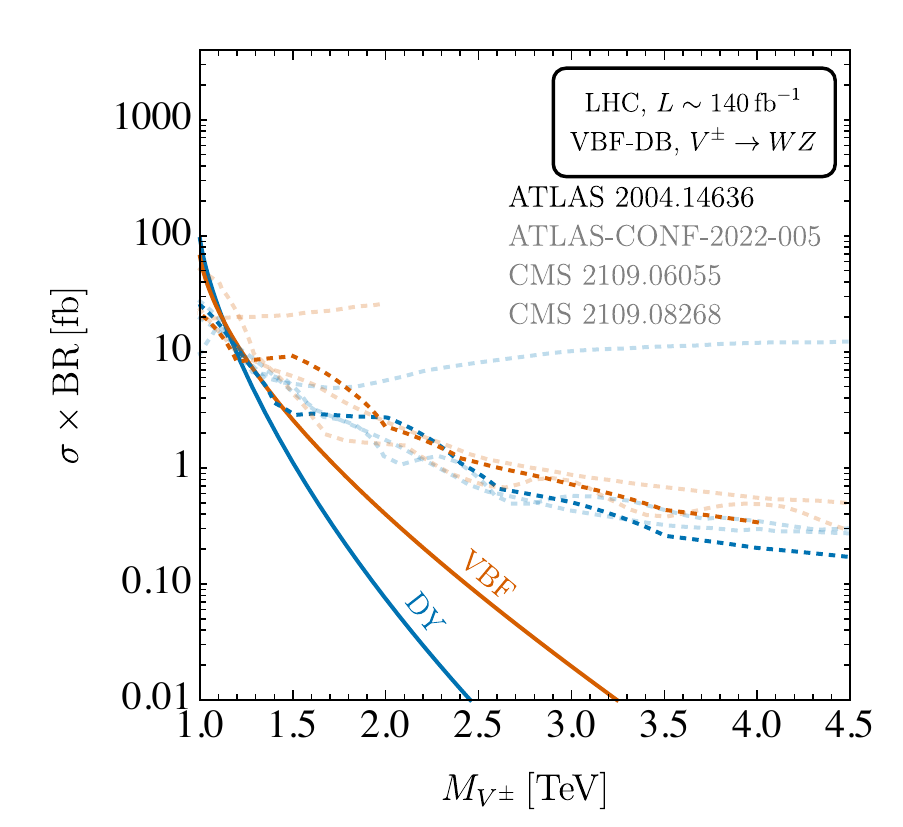}
    \includegraphics[width=0.49\textwidth]{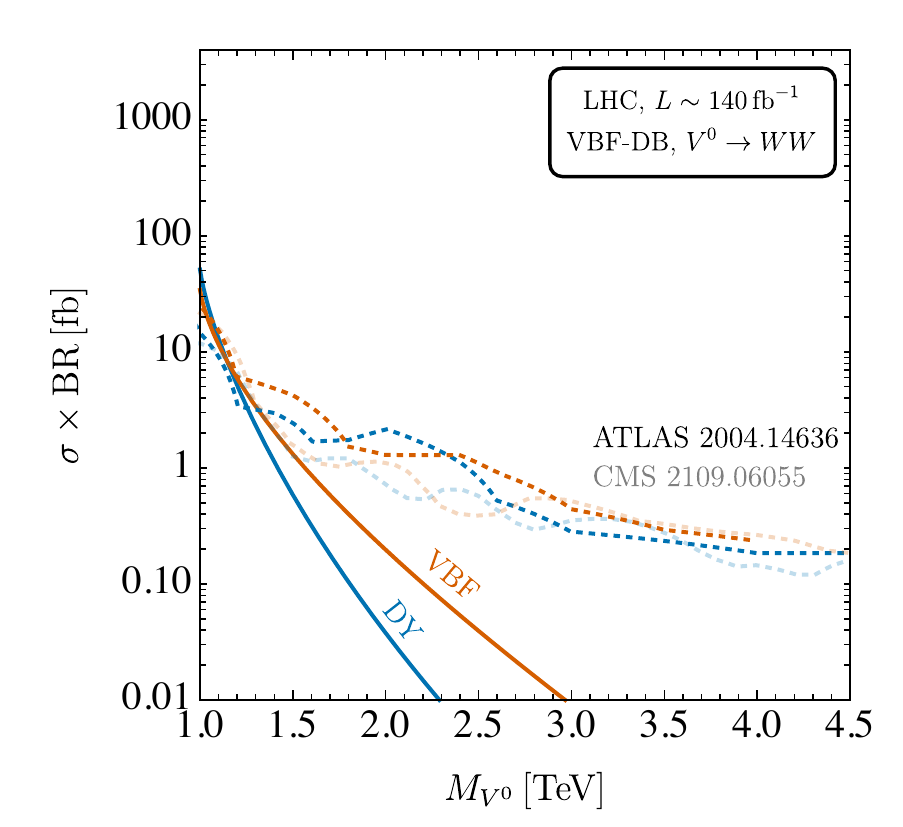}
    \caption{
        The DY (blue) and VBF (red) production cross-sections times branching ratio in $WZ$ (left) and $WW$ (right) for the VBF-DB benchmark. The most stringent LHC limits from~\cite{ATLAS:2020fry} are shown in bright dotted blue and red, while similar but less constraining searches are shown in the background~\cite{ATLAS:2022jho,CMS:2021klu,CMS:2021itu}.
    }
    \label{fig:cs-br-zh}
\end{figure}

\Cref{tab:expsearches} lists the most recent experimental searches performed by ATLAS and CMS with an integrated luminosity of $\sim 140\,\text{fb}^{-1}$ looking for di-boson and di-lepton signatures. In \cref{fig:cs-br-zh} we show $\sigma \times \text{BR}$ as a function of the HVT mass for decays into $WZ$ (left) and $WW$ (right). The experimental DY and VBF di-boson searches are shown in dotted blue and red, respectively. The solid blue and red curves depict $\sigma \times \text{BR}$ via DY and VBF for the VBF-DB benchmark. Currently, both DY and VBF production can probe masses up to $1.2\,$TeV in $WZ$ final states. We see that while $\sigma \times \text{BR}$ is comparable for DY and VBF for masses around $1\,$TeV, for larger masses the VBF contribution becomes significantly larger than the DY one. For neutral $WW$ final states, the current mass reach of both DY and VBF analyses is currently at $1.2\,$TeV. VBF searches in $Wh$ final states have not yet been performed while $Zh$ searches are currently slightly weaker than $WW$ final states. Despite similar current limits, we see that future VBF analyses are expected to have a better sensitivity than future DY analyses in this benchmark region of HVT parameter space.

\begin{figure}
    \centering
    \rotatebox{90}{\phantom{XXXXXXXll}
    \scalebox{0.7}{$c_\ell/g_V$}}
    \includegraphics[width=0.31\textwidth]{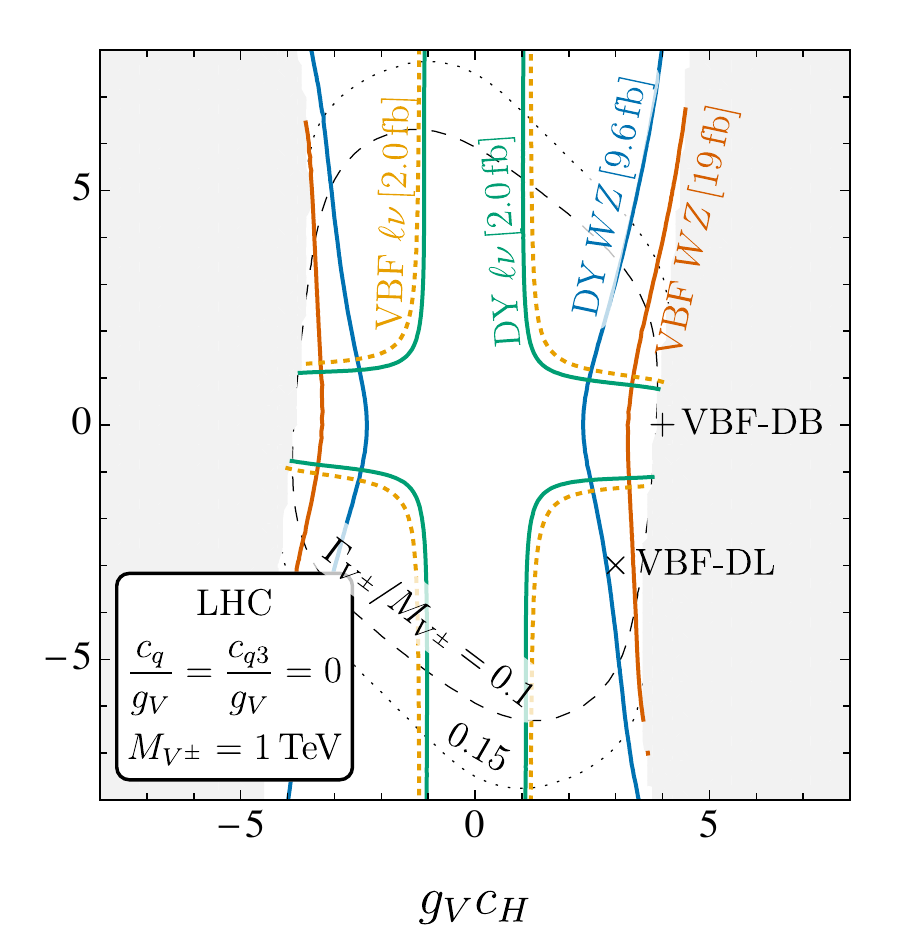}
    \includegraphics[width=0.31\textwidth]{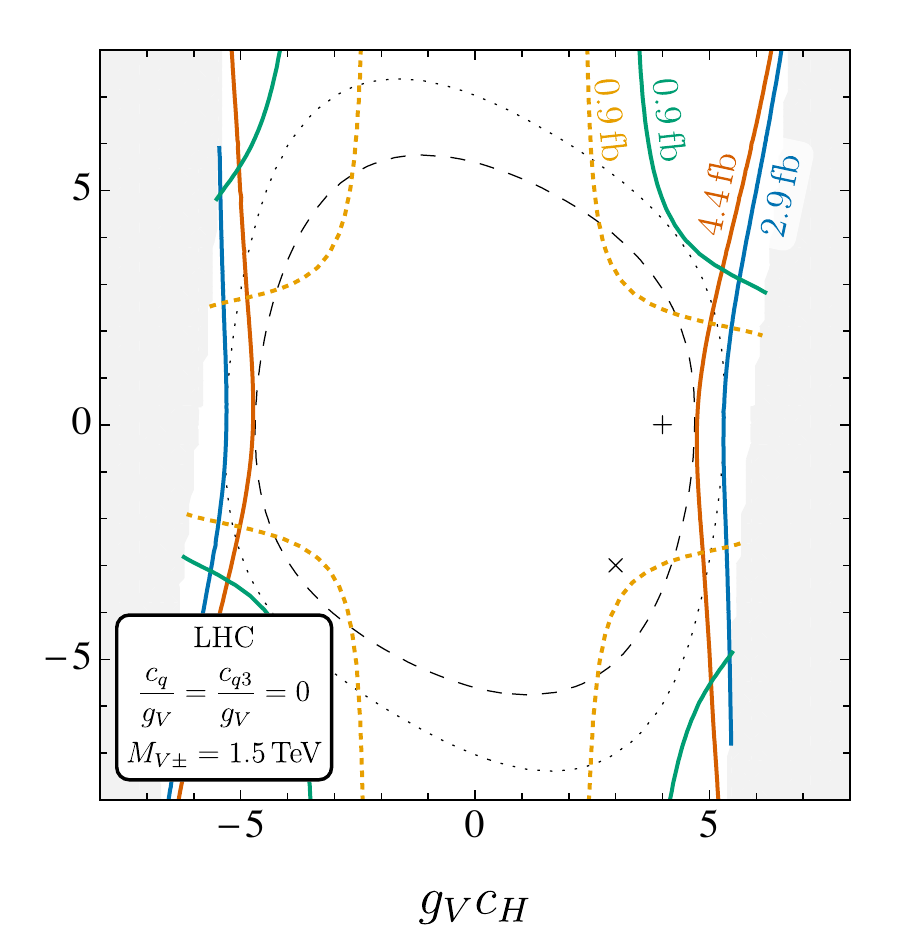}
    \includegraphics[width=0.31\textwidth]{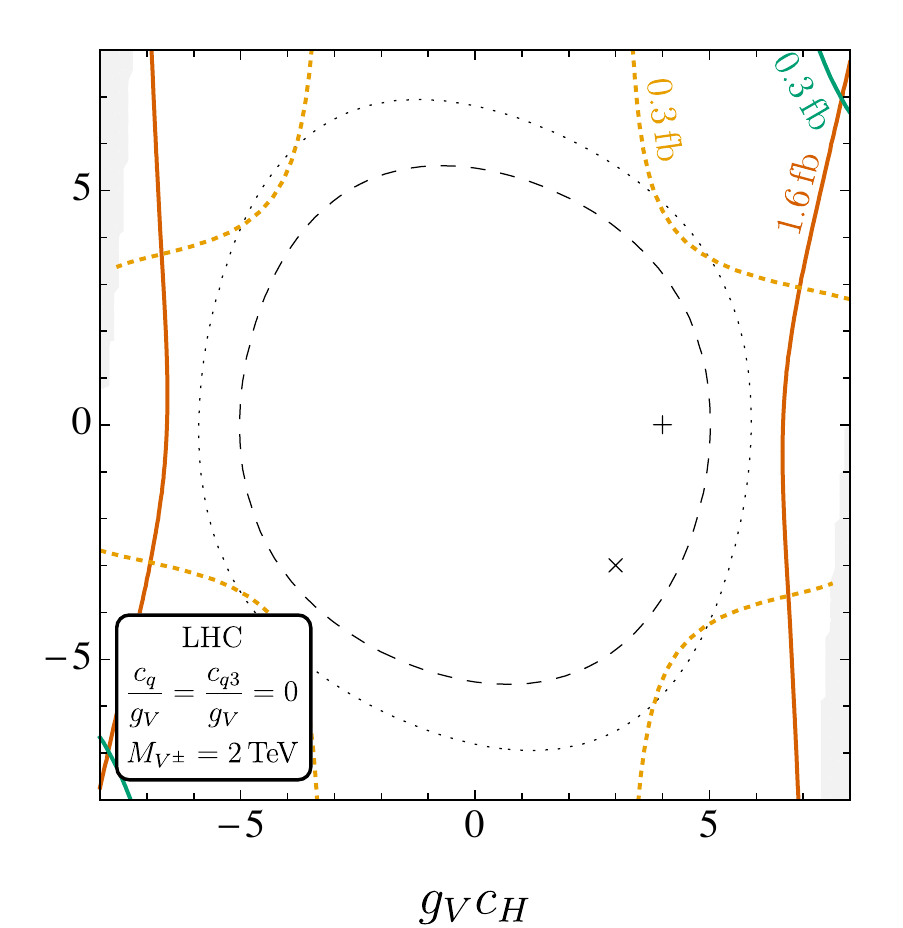}\\
    \rotatebox{90}{\phantom{XXXXXXXll}
    \scalebox{0.7}{$c_\ell/g_V$}}
    \includegraphics[width=0.31\textwidth]{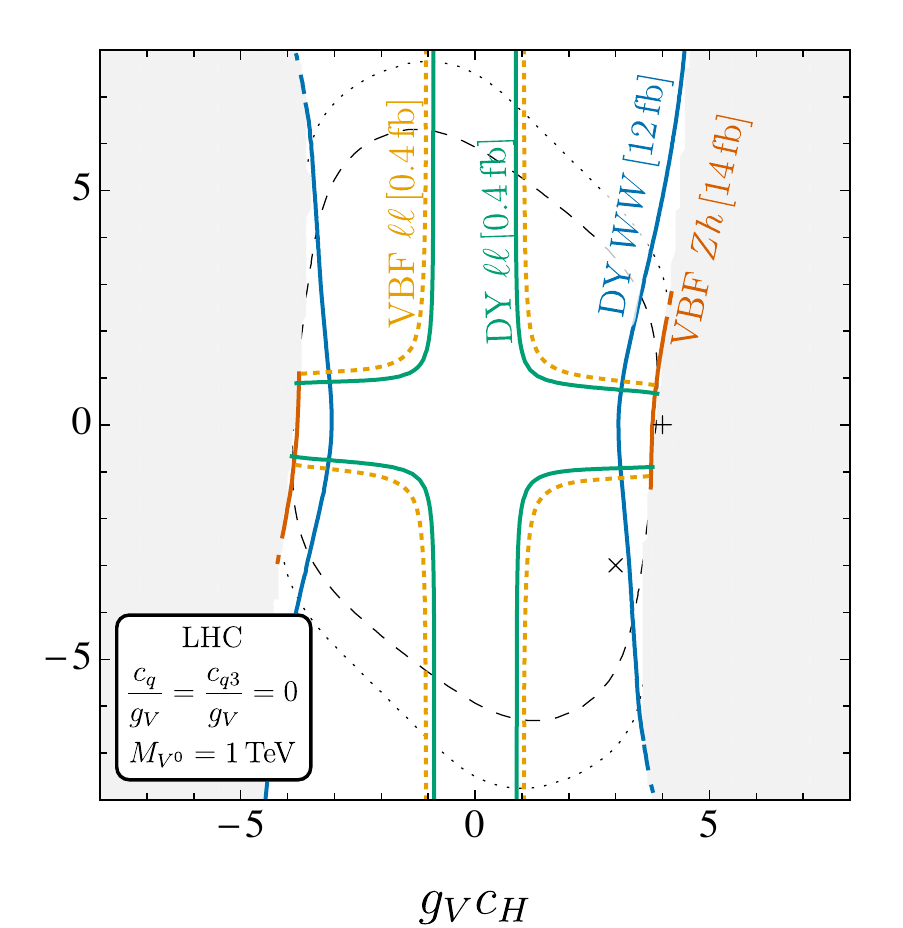}
    \includegraphics[width=0.31\textwidth]{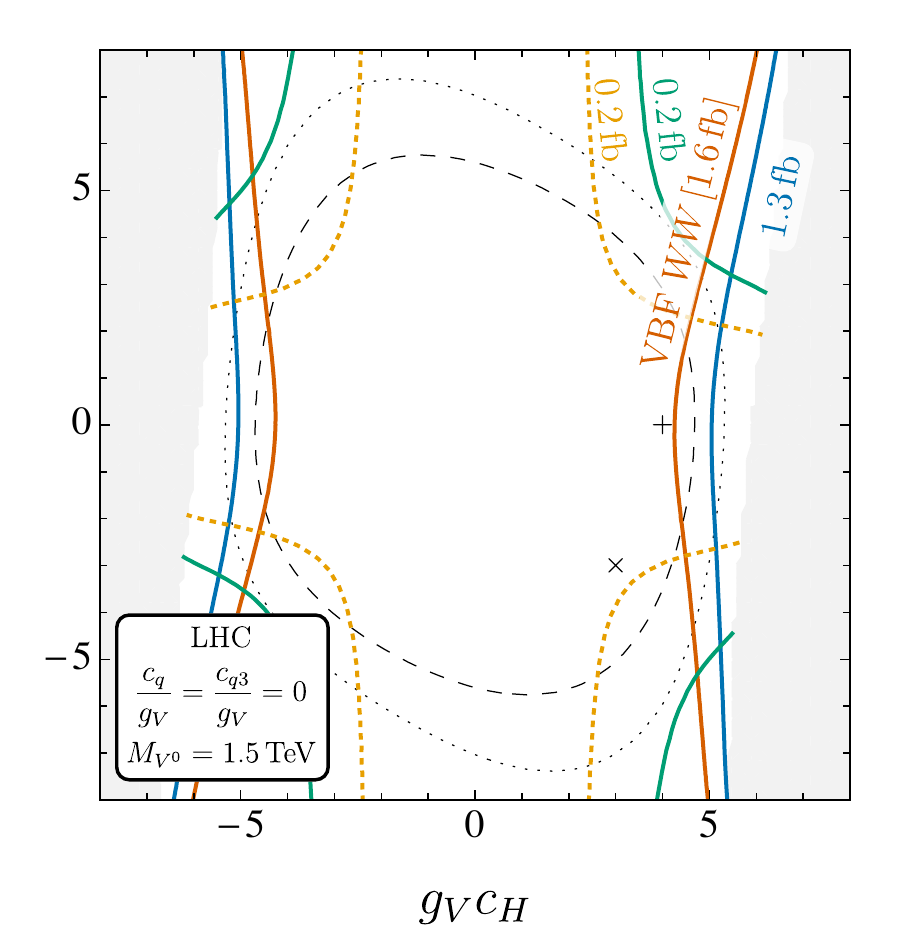}
    \includegraphics[width=0.31\textwidth]{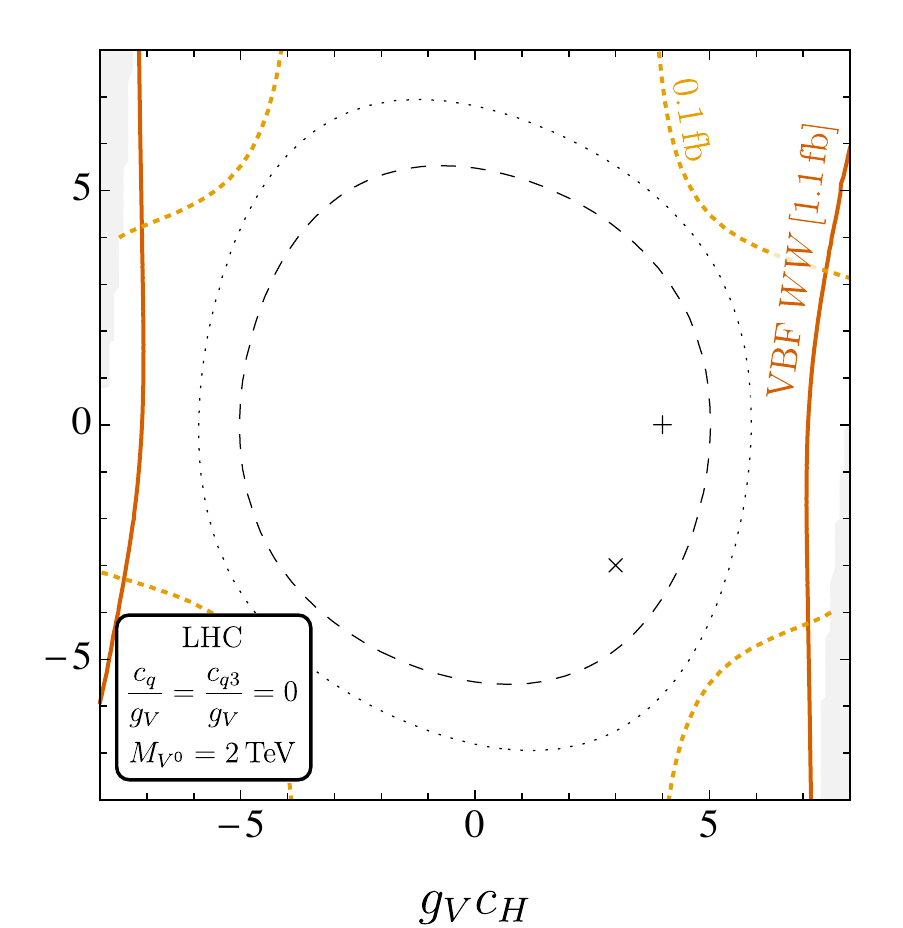}
    \caption{Current limits in the $g_V c_H - c_\ell/g_V$ parameter plane for charged (top) and neutral (bottom) HVT resonances. Blue and red lines show the limits set by the most constraining DY and VBF searches in di-boson final states~\cite{ATLAS:2020fry,CMS:2021klu,ATLAS:2022jho}. The solid green and dotted yellow lines shows the most constraining lepton-neutrino~\cite{CMS:2022yjm,ATLAS:2019lsy} (top) and di-lepton limit~\cite{ATLAS:2019erb,CMS:2021ctt} (bottom) for DY (solid green) and VBF (dotted yellow) assuming the same sensitivity as for DY.  The dashed and dotted black lines show contours of constant $\Gamma_\text{tot}/M_V = 0.1$ and $0.15$. The $+$ and $\times$ symbols indicate the VBF-DB and VBF-DL benchmark parameter points, respectively. The grey regions are theoretically excluded in the simplified model.
    }
    \label{fig:limits}
\end{figure}

\Cref{fig:limits} shows the experimental limits for a charged (top) and neutral (bottom) HVT resonance in the $g_V c_H - c_\ell / g_V$ parameter plane for masses of $1, 1.5$ and $2\,$TeV (left, centre, right). The blue and red lines depict the most sensitive experimental limits in the di-boson final state in the DY and VBF channels~\cite{ATLAS:2020fry,CMS:2021klu,ATLAS:2022jho}, respectively. For the charged resonance there are no $Wh$ searches so $WZ$ always provides the best di-boson limit. For the neutral one, DY $WW$ always provides the best DY limit while VBF $Zh$ is best at 1\,TeV and VBF $WW$ is stronger at 1.5 and 2\,TeV. While resonance production via DY leads to stronger constraints than VBF at $1\,$TeV, VBF production outperforms DY production at $1.5\,$TeV and is the only sensitive production mode at $2\,$TeV. The other di-boson searches lead to similar but slightly weaker constraints.
The solid green line depicts the reach of lepton-neutrino~\cite{CMS:2022yjm,ATLAS:2019lsy} and di-lepton~\cite{ATLAS:2019erb,CMS:2021ctt} searches in DY. A di-lepton search in the VBF channel has not been performed yet. Given the similar reach of current di-boson searches in the DY and VBF channels, we make the assumption that the di-lepton search using VBF production will be similar to the current limit using DY and include that prediction as a yellow dotted line. Analogously to di-boson final states, we see that DY production is most sensitive at $1\,$TeV but that VBF production is expected to outperform DY searches at $1.5\,$TeV. For a resonance mass of $2\,$TeV VBF is again expected to be the only sensitive production mode.
The dashed and dotted black lines show contours of constant $\Gamma_\text{tot}/M_V = 10\%$ and $15\%$. The grey regions are theoretically excluded in the HVT model.

\subsection{Projected Limits at the HL-LHC}
\label{subsec:hl-lhc-limits}

\begin{figure}
    \centering
    \includegraphics[width=0.45\textwidth]{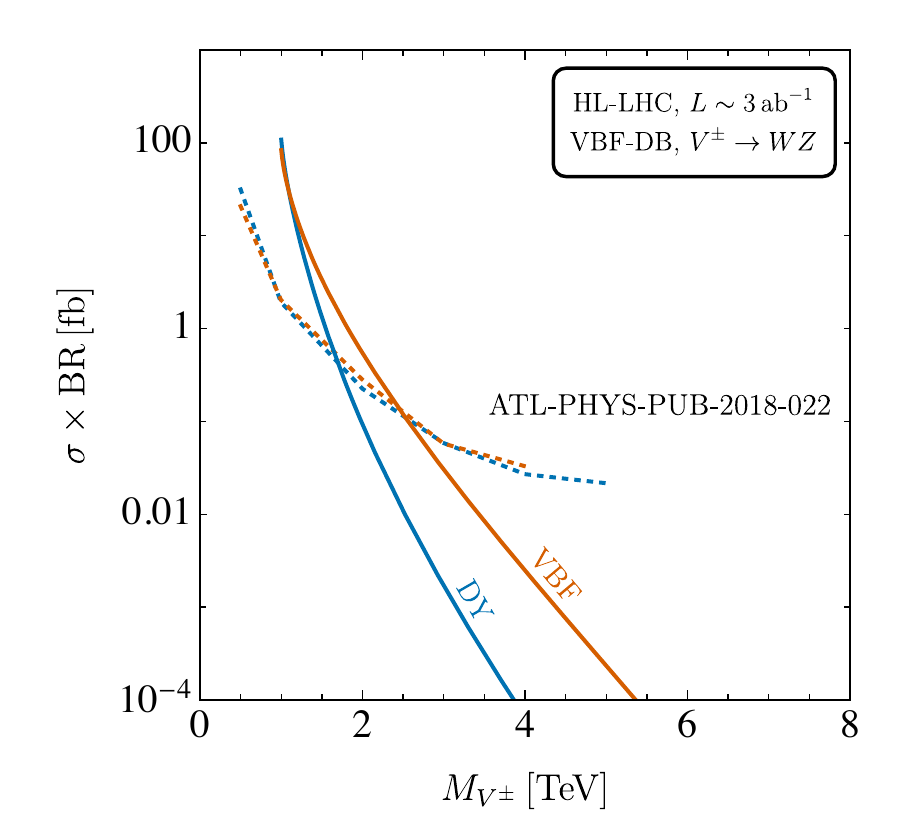}
    \includegraphics[width=0.45\textwidth]{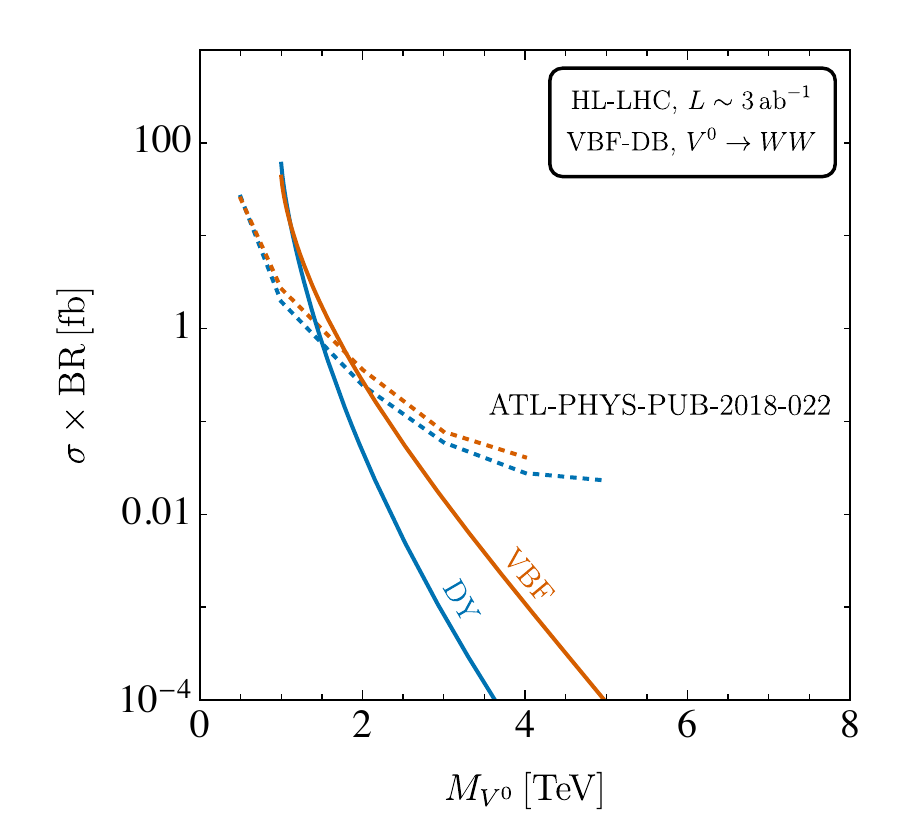}\\
    \includegraphics[width=0.45\textwidth]{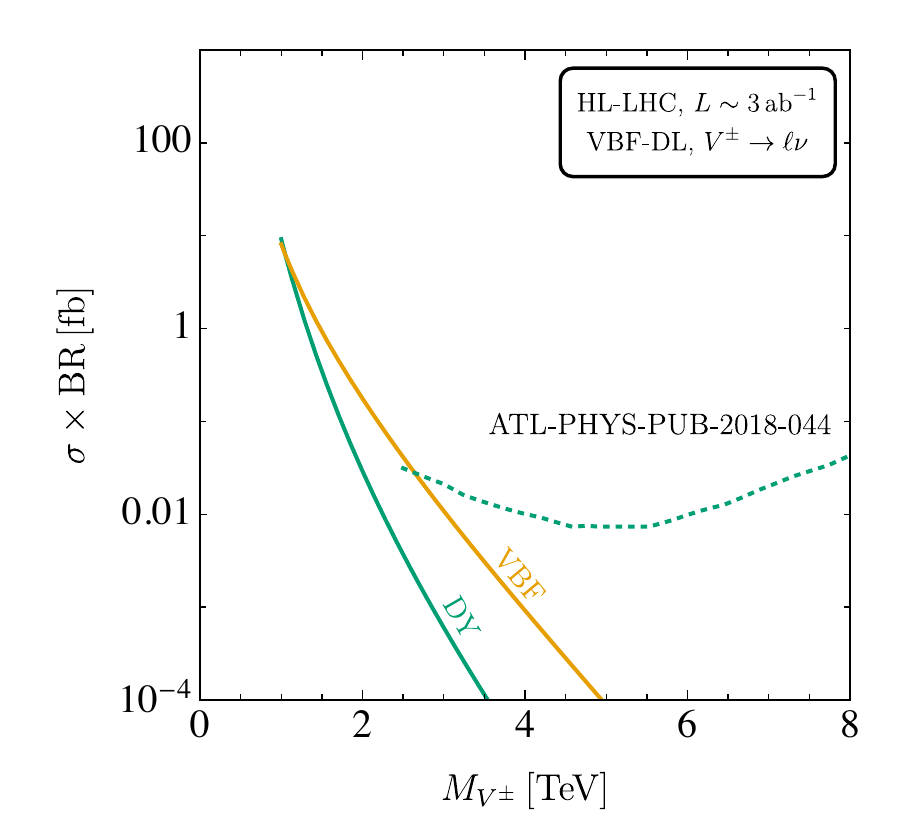}
    \includegraphics[width=0.45\textwidth]{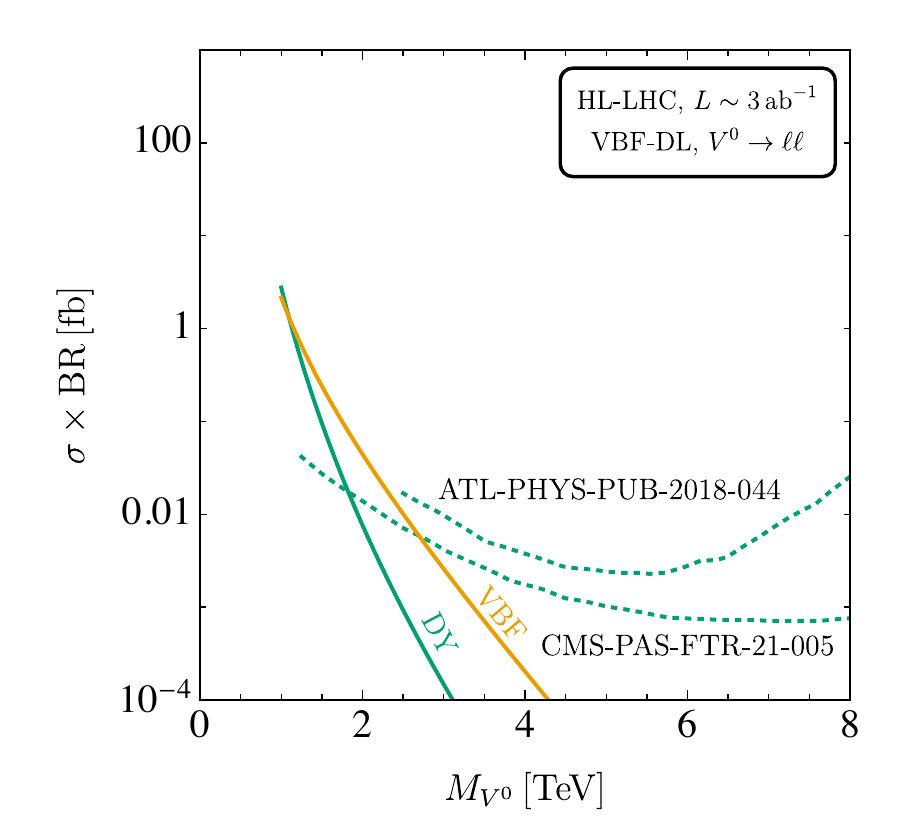}
    \caption{
        (Top) Production cross-section times branching ratio in $WZ$ (left) and $WW$ (right) for the VBF di-boson benchmark for DY and VBF production in blue and red, respectively. Projections for the HL-LHC limits at $14\,$TeV with $3\,$ab$^{-1}$~\cite{ATLAS:2018ocj} are shown in dotted blue (DY) and dotted red (VBF). 
        (Bottom)
        Production cross-section times branching ratio in $\ell\nu$ (left) and $\ell\ell$ (right) for the VBF di-lepton benchmark for DY production and VBF in green and yellow, respectively. Projections of HL-LHC limits at $14\,$TeV with $3\,$ab$^{-1}$~\cite{ATLAS:2018tvr,CMS:2022gho} are shown in dotted green (DY).
    }
    \label{fig:hl-lhc-cs-br-mass}
\end{figure}

In the future, the LHC is well placed to further leverage the dominant VBF production mode present in certain regions of the HVT parameter space. \Cref{fig:hl-lhc-cs-br-mass} shows exclusion projections for $\sigma \times \text{BR}$ at the HL-LHC running at $14\,$TeV with an integrated luminosity of $3\,$ab$^{-1}$, compared to the HVT $\sigma \times \text{BR}$ into di-bosons (top) for the VBF di-boson benchmark parameters and into di-leptons (bottom) for the VBF di-lepton benchmark, for both DY and VBF production. On the left we show projections for the charged component of the HVT and on the right for the neutral component. 

For di-boson final states, exclusion projections for DY and VBF have been derived in ref.~\cite{ATLAS:2018ocj} and are shown here in dotted blue for DY and in dotted red for VBF. We see the significantly higher mass reach in VBF studies with respect to DY: while DY production will be able to probe charged resonance masses up to $1.7\,$TeV, the VBF topology can push the mass limit up to $2.5\,$TeV. Analogously, a neutral resonance search is expected to be sensitive up to masses of $1.5\,$TeV in DY production and up to $1.9\,$TeV using VBF.

HL-LHC exclusion projections for lepton neutrino (bottom left) and di-lepton (bottom right) final states have only been derived for DY production \cite{ATLAS:2018tvr,CMS:2022gho}, shown here in dotted green. Extrapolating the lepton-neutrino limit to smaller masses, we would expect a mass reach of roughly $2\,$TeV. The neutral di-lepton final states will lead to a similar reach in DY production. Although the experimental collaborations have not derived projections for di-lepton final states at the HL-LHC using VBF production,\footnote{VBF projections have been discussed in specific channels at various future colliders in refs.~\cite{Mohan:2015doa, Cavaliere:2018zcf, Kim:2021vxd}. } we expect the VBF limit to have a similar sensitivity to DY production given that DY and VBF production lead to similar exclusion bounds in current di-boson searches. With this assumption, VBF limits would be expected to probe masses up to $2.6\,$TeV for both charged and neutral di-lepton final states.

At a future $100\,$TeV collider, projections for the expected sensitivity in DY production were partially derived in ref.~\cite{Thamm:2015zwa} using a dedicated extrapolation procedure which is based on assumptions that are not straightforwardly satisfied for VBF production. Similar projections for the VBF sensitivity at $100\,$TeV would require a dedicated analysis which we postpone to future work.

\section{Conclusions}
\label{sec:conclusions}

In this paper we have discussed the role of vector boson fusion for the production of heavy vector triplets. We analysed the production cross-section of a heavy vector in terms of a numerical factor, the parton luminosities and the partial widths. We pointed out that the presence of VBF production generically leads to a comparable rate of DY production, even with vanishing coupling to light quarks, due to mixing effects. While the parton luminosities decrease with increasing HVT mass, the importance of the partial widths into di-bosons increases with larger resonance masses in this region of parameter space, leading to a delicate interplay which favours DY production at lower masses and VBF at higher masses. We have thus demonstrated that vector boson fusion is a competitive production mode for heavy vector triplets in certain regions of parameter space.

The most interesting decay channels for heavy vector triplets produced predominantly in VBF are decay modes into di-bosons and di-leptons. We defined two benchmark parameter points, the VBF di-boson and di-lepton benchmarks, which provide competitive resonance production via vector boson fusion and allow for decays into di-bosons only (VBF di-boson benchmark) or decays into di-leptons (VBF di-lepton benchmark). In both cases, while DY di-boson analyses set the most stringent constraints for resonance masses of $1\,$TeV, we have shown that VBF production becomes more constraining at $1.5\,$TeV and the only sensitive search strategy at $2\,$TeV. Current LHC and projected HL-LHC limits exemplify the higher mass reach of VBF searches with respect to DY analyses and thus highlight the importance of VBF production in these regions of the HVT parameter space.

\section*{Acknowledgements}
\label{sec:acknowledgements}

MJB would like to acknowledge the support of the Australian Research Council through the ARC Centre of Excellence for Dark Matter Particle Physics, CE200100008.
MJB and AT would like to thank the IPPP at Durham University for kind hospitality during a significant portion of the work and the Mainz Institute for Theoretical Physics (MITP) of the Cluster of Excellence PRISMA+ (Project ID 39083149) for its hospitality and partial support during the completion of this work.
The work of RT was partly
supported by the Italian Ministry of Research (MIUR) under the PRIN grant 20172LNEEZ.

\bibliographystyle{JHEP}
\bibliography{refs}

\providecommand{\href}[2]{#2}\begingroup\raggedright\begin{thebibliography}{10}

\bibitem{Barger:1980ix}
V.~D. Barger, W.-Y. Keung, and E.~Ma, {\it {A Gauge Model With Light $W$ and
  $Z$ Bosons}},  {\em Phys. Rev. D} {\bf 22} (1980) 727.

\bibitem{Hewett:1988xc}
J.~L. Hewett and T.~G. Rizzo, {\it {Low-Energy Phenomenology of Superstring
  Inspired E(6) Models}},  {\em Phys. Rept.} {\bf 183} (1989) 193.

\bibitem{Cvetic:1995zs}
M.~Cvetic and S.~Godfrey, {\em {Discovery and identification of extra gauge
  bosons}}, pp.~383--415.
\newblock 3, 1995.
\newblock \href{http://arxiv.org/abs/hep-ph/9504216}{{\tt hep-ph/9504216}}.

\bibitem{Rizzo:2006nw}
T.~G. Rizzo, {\it {$Z^\prime$ phenomenology and the LHC}},  in {\em
  {Theoretical Advanced Study Institute in Elementary Particle Physics}:
  {Exploring New Frontiers Using Colliders and Neutrinos}}, pp.~537--575, 10,
  2006.
\newblock \href{http://arxiv.org/abs/hep-ph/0610104}{{\tt hep-ph/0610104}}.

\bibitem{Langacker:2008yv}
P.~Langacker, {\it {The Physics of Heavy $Z^\prime$ Gauge Bosons}},  {\em Rev.
  Mod. Phys.} {\bf 81} (2009) 1199--1228,
  [\href{http://arxiv.org/abs/0801.1345}{{\tt arXiv:0801.1345}}].

\bibitem{Salvioni:2009mt}
E.~Salvioni, G.~Villadoro, and F.~Zwirner, {\it {Minimal Z-prime models:
  Present bounds and early LHC reach}},  {\em JHEP} {\bf 11} (2009) 068,
  [\href{http://arxiv.org/abs/0909.1320}{{\tt arXiv:0909.1320}}].

\bibitem{Salvioni:2009jp}
E.~Salvioni, A.~Strumia, G.~Villadoro, and F.~Zwirner, {\it {Non-universal
  minimal Z' models: present bounds and early LHC reach}},  {\em JHEP} {\bf 03}
  (2010) 010, [\href{http://arxiv.org/abs/0911.1450}{{\tt arXiv:0911.1450}}].

\bibitem{Chanowitz:2011ew}
M.~S. Chanowitz, {\it {A Heavy little Higgs and a light Z' under the radar}},
  {\em Phys. Rev. D} {\bf 84} (2011) 035014,
  [\href{http://arxiv.org/abs/1102.3672}{{\tt arXiv:1102.3672}}].

\bibitem{Langacker:1989xa}
P.~Langacker and S.~U. Sankar, {\it {Bounds on the Mass of W(R) and the
  W(L)-W(R) Mixing Angle xi in General SU(2)-L x SU(2)-R x U(1) Models}},  {\em
  Phys. Rev. D} {\bf 40} (1989) 1569--1585.

\bibitem{Sullivan:2002jt}
Z.~Sullivan, {\it {Fully Differential $W^{\prime}$ Production and Decay at
  Next-to-Leading Order in QCD}},  {\em Phys. Rev. D} {\bf 66} (2002) 075011,
  [\href{http://arxiv.org/abs/hep-ph/0207290}{{\tt hep-ph/0207290}}].

\bibitem{Grojean:2011vu}
C.~Grojean, E.~Salvioni, and R.~Torre, {\it {A weakly constrained W' at the
  early LHC}},  {\em JHEP} {\bf 07} (2011) 002,
  [\href{http://arxiv.org/abs/1103.2761}{{\tt arXiv:1103.2761}}].

\bibitem{Schmaltz:2010xr}
M.~Schmaltz and C.~Spethmann, {\it {Two Simple W' Models for the Early LHC}},
  {\em JHEP} {\bf 07} (2011) 046, [\href{http://arxiv.org/abs/1011.5918}{{\tt
  arXiv:1011.5918}}].

\bibitem{Frank:2010cj}
M.~Frank, A.~Hayreter, and I.~Turan, {\it {Production and Decays of $W_R$
  bosons at the LHC}},  {\em Phys. Rev. D} {\bf 83} (2011) 035001,
  [\href{http://arxiv.org/abs/1010.5809}{{\tt arXiv:1010.5809}}].

\bibitem{Agashe:2007ki}
K.~Agashe, H.~Davoudiasl, S.~Gopalakrishna, T.~Han, G.-Y. Huang, G.~Perez,
  Z.-G. Si, and A.~Soni, {\it {LHC Signals for Warped Electroweak Neutral Gauge
  Bosons}},  {\em Phys. Rev. D} {\bf 76} (2007) 115015,
  [\href{http://arxiv.org/abs/0709.0007}{{\tt arXiv:0709.0007}}].

\bibitem{Agashe:2008jb}
K.~Agashe, S.~Gopalakrishna, T.~Han, G.-Y. Huang, and A.~Soni, {\it {LHC
  Signals for Warped Electroweak Charged Gauge Bosons}},  {\em Phys. Rev. D}
  {\bf 80} (2009) 075007, [\href{http://arxiv.org/abs/0810.1497}{{\tt
  arXiv:0810.1497}}].

\bibitem{Agashe:2009bb}
K.~Agashe, A.~Azatov, T.~Han, Y.~Li, Z.-G. Si, and L.~Zhu, {\it {LHC Signals
  for Coset Electroweak Gauge Bosons in Warped/Composite PGB Higgs Models}},
  {\em Phys. Rev. D} {\bf 81} (2010) 096002,
  [\href{http://arxiv.org/abs/0911.0059}{{\tt arXiv:0911.0059}}].

\bibitem{Contino:2011np}
R.~Contino, D.~Marzocca, D.~Pappadopulo, and R.~Rattazzi, {\it {On the effect
  of resonances in composite Higgs phenomenology}},  {\em JHEP} {\bf 10} (2011)
  081, [\href{http://arxiv.org/abs/1109.1570}{{\tt arXiv:1109.1570}}].

\bibitem{Bellazzini:2012tv}
B.~Bellazzini, C.~Csaki, J.~Hubisz, J.~Serra, and J.~Terning, {\it {Composite
  Higgs Sketch}},  {\em JHEP} {\bf 11} (2012) 003,
  [\href{http://arxiv.org/abs/1205.4032}{{\tt arXiv:1205.4032}}].

\bibitem{Accomando:2012yg}
E.~Accomando, L.~Fedeli, S.~Moretti, S.~De~Curtis, and D.~Dominici, {\it
  {Charged di-boson production at the LHC in a 4-site model with a composite
  Higgs boson}},  {\em Phys. Rev. D} {\bf 86} (2012) 115006,
  [\href{http://arxiv.org/abs/1208.0268}{{\tt arXiv:1208.0268}}].

\bibitem{CarcamoHernandez:2013ydh}
A.~E. Carcamo~Hernandez, C.~O. Dib, and A.~R. Zerwekh, {\it {The Effect of
  Composite Resonances on Higgs decay into two photons}},  {\em Eur. Phys. J.
  C} {\bf 74} (2014) 2822, [\href{http://arxiv.org/abs/1304.0286}{{\tt
  arXiv:1304.0286}}].

\bibitem{Low:2015uha}
M.~Low, A.~Tesi, and L.-T. Wang, {\it {Composite spin-1 resonances at the
  LHC}},  {\em Phys. Rev. D} {\bf 92} (2015), no.~8 085019,
  [\href{http://arxiv.org/abs/1507.07557}{{\tt arXiv:1507.07557}}].

\bibitem{Accomando:2016mvz}
E.~Accomando, D.~Barducci, S.~De~Curtis, J.~Fiaschi, S.~Moretti, and C.~H.
  Shepherd-Themistocleous, {\it {Drell-Yan production of multi Z$^{'}$-bosons
  at the LHC within Non-Universal ED and 4D Composite Higgs Models}},  {\em
  JHEP} {\bf 07} (2016) 068, [\href{http://arxiv.org/abs/1602.05438}{{\tt
  arXiv:1602.05438}}].

\bibitem{Liu:2018hum}
D.~Liu, L.-T. Wang, and K.-P. Xie, {\it {Prospects of searching for composite
  resonances at the LHC and beyond}},  {\em JHEP} {\bf 01} (2019) 157,
  [\href{http://arxiv.org/abs/1810.08954}{{\tt arXiv:1810.08954}}].

\bibitem{Capdevilla:2019zbx}
R.~M. Capdevilla, R.~Harnik, and A.~Martin, {\it {The radiation valley and
  exotic resonances in $W\gamma$ production at the LHC}},  {\em JHEP} {\bf 03}
  (2020) 117, [\href{http://arxiv.org/abs/1912.08234}{{\tt arXiv:1912.08234}}].

\bibitem{deBlas:2012tc}
J.~{de Blas}, J.~M. Lizana, and M.~Perez-Victoria, {\it {Combining searches of
  $Z'$ and $W'$ bosons}},  {\em JHEP} {\bf 01} (2013) 166,
  [\href{http://arxiv.org/abs/1211.2229}{{\tt arXiv:1211.2229}}].
  [\href{http://inspirehep.net/record/1201947}{Inspire}].

\bibitem{Pappadopulo:2014tg}
D.~Pappadopulo, A.~Thamm, A.~Wulzer, and R.~Torre, {\it {Heavy Vector Triplets:
  Bridging Theory and Data}},  \href{http://arxiv.org/abs/1402.4431}{{\tt
  arXiv:1402.4431}}. [\href{http://inspirehep.net/record/1281686}{Inspire}].

\bibitem{Chivukula:2017lyk}
R.~S. Chivukula, P.~Ittisamai, K.~Mohan, and E.~H. Simmons, {\it {Broadening
  the Reach of Simplified Limits on Resonances at the LHC}},  {\em Phys. Rev.
  D} {\bf 96} (2017), no.~5 055043,
  [\href{http://arxiv.org/abs/1707.01080}{{\tt arXiv:1707.01080}}].

\bibitem{Chivukula:2021foa}
R.~S. Chivukula, P.~Ittisamai, J.~Osborne, and E.~H. Simmons, {\it {Narrow
  Resonances Revisited -- Simplifying Multidimensional Constraints}},  {\em
  Phys. Rev. D} {\bf 103} (2021), no.~9 095008,
  [\href{http://arxiv.org/abs/2103.06283}{{\tt arXiv:2103.06283}}].

\bibitem{HVTGitHub}
R.~Torre, {\it {HVT Tools}},  2022.
\newblock \href{httpshttps://github.com/riccardotorre/HVT_tools}{GitHub}.

\bibitem{Mohan:2015doa}
K.~Mohan and N.~Vignaroli, {\it {Vector resonances in weak-boson-fusion at
  future pp colliders}},  {\em JHEP} {\bf 10} (2015) 031,
  [\href{http://arxiv.org/abs/1507.03940}{{\tt arXiv:1507.03940}}].

\bibitem{Florez:2016uob}
A.~Fl\'orez, A.~Gurrola, W.~Johns, Y.~D. Oh, P.~Sheldon, D.~Teague, and
  T.~Weiler, {\it {Searching for New Heavy Neutral Gauge Bosons using Vector
  Boson Fusion Processes at the LHC}},  {\em Phys. Lett. B} {\bf 767} (2017)
  126--132, [\href{http://arxiv.org/abs/1609.09765}{{\tt arXiv:1609.09765}}].

\bibitem{Cavaliere:2018zcf}
V.~Cavaliere, R.~Les, T.~Nitta, and K.~Terashi, {\it {HE-LHC prospects for
  diboson resonance searches and electroweak WW/WZ production via vector boson
  scattering in the semi-leptonic final states}},
  \href{http://arxiv.org/abs/1812.00841}{{\tt arXiv:1812.00841}}.

\bibitem{Kim:2021vxd}
D.~Kim, Y.~Oh, B.~Tae, and J.~Lee, {\it {Prospect of the search for ${\mathrm
  {W}}^{\prime }$ in vector boson fusion at the high-luminosity large hadron
  collider}},  {\em J. Korean Phys. Soc.} {\bf 78} (2021), no.~3 182--187.

\bibitem{ATLAS:2022jho}
{\bf ATLAS} Collaboration, {\it {Search for Resonant $WZ \rightarrow \ell\nu
  \ell^{\prime}\ell^{\prime}$ Production in Proton-Proton Collisions at
  $\mathbf{\sqrt{s} = 13}$ TeV with the ATLAS Detector}},  ATLAS-CONF-2022-005.

\bibitem{ATLAS:2020fry}
{\bf ATLAS} Collaboration, G.~Aad et~al., {\it {Search for heavy diboson
  resonances in semileptonic final states in pp collisions at $\sqrt{s}=13$ TeV
  with the ATLAS detector}},  {\em Eur. Phys. J. C} {\bf 80} (2020), no.~12
  1165, [\href{http://arxiv.org/abs/2004.14636}{{\tt arXiv:2004.14636}}].

\bibitem{ATLAS:2018sbw}
{\bf ATLAS} Collaboration, M.~Aaboud et~al., {\it {Combination of searches for
  heavy resonances decaying into bosonic and leptonic final states using 36
  fb$^{-1}$ of proton-proton collision data at $\sqrt{s} = 13$ TeV with the
  ATLAS detector}},  {\em Phys. Rev. D} {\bf 98} (2018), no.~5 052008,
  [\href{http://arxiv.org/abs/1808.02380}{{\tt arXiv:1808.02380}}].

\bibitem{ATLAS:2018iui}
{\bf ATLAS} Collaboration, M.~Aaboud et~al., {\it {Search for resonant $WZ$
  production in the fully leptonic final state in proton-proton collisions at
  $\sqrt{s} = 13$ TeV with the ATLAS detector}},  {\em Phys. Lett. B} {\bf 787}
  (2018) 68--88, [\href{http://arxiv.org/abs/1806.01532}{{\tt
  arXiv:1806.01532}}].

\bibitem{ATLAS:2018ocj}
{\bf ATLAS} Collaboration, {\it {HL-LHC prospects for diboson resonance
  searches and electroweak vector boson scattering in the $WW/WZ\to\ell\nu qq$
  final state}},  ATL-PHYS-PUB-2018-022.

\bibitem{ATLAS:2017uhp}
{\bf ATLAS} Collaboration, M.~Aaboud et~al., {\it {Search for heavy resonances
  decaying into $WW$ in the $e\nu\mu\nu$ final state in $pp$ collisions at
  $\sqrt{s}=13$ TeV with the ATLAS detector}},  {\em Eur. Phys. J. C} {\bf 78}
  (2018), no.~1 24, [\href{http://arxiv.org/abs/1710.01123}{{\tt
  arXiv:1710.01123}}].

\bibitem{ATLAS:2017otj}
{\bf ATLAS} Collaboration, M.~Aaboud et~al., {\it {Searches for heavy $ZZ$ and
  $ZW$ resonances in the $\ell\ell qq$ and $\nu\nu qq$ final states in $pp$
  collisions at $\sqrt{s}=13$ TeV with the ATLAS detector}},  {\em JHEP} {\bf
  03} (2018) 009, [\href{http://arxiv.org/abs/1708.09638}{{\tt
  arXiv:1708.09638}}].

\bibitem{ATLAS:2017jag}
{\bf ATLAS} Collaboration, M.~Aaboud et~al., {\it {Search for $WW/WZ$ resonance
  production in $\ell \nu qq$ final states in $pp$ collisions at $\sqrt{s} =$
  13 TeV with the ATLAS detector}},  {\em JHEP} {\bf 03} (2018) 042,
  [\href{http://arxiv.org/abs/1710.07235}{{\tt arXiv:1710.07235}}].

\bibitem{CMS:2022shx}
{\bf CMS} Collaboration, {\it {Search for new heavy resonances decaying to WW,
  WZ, ZZ, WH, or ZH boson pairs in the all-jets final state in proton-proton
  collisions at $\sqrt{s}=13~\mathrm{TeV}$}},  CMS-PAS-B2G-20-009.

\bibitem{CMS:2021klu}
{\bf CMS} Collaboration, A.~Tumasyan et~al., {\it {Search for heavy resonances
  decaying to WW, WZ, or WH boson pairs in the lepton plus merged jet final
  state in proton-proton collisions at $\sqrt{s}$ = 13 TeV}},  {\em Phys. Rev.
  D} {\bf 105} (2022), no.~3 032008,
  [\href{http://arxiv.org/abs/2109.06055}{{\tt arXiv:2109.06055}}].

\bibitem{CMS:2021itu}
{\bf CMS} Collaboration, A.~Tumasyan et~al., {\it {Search for heavy resonances
  decaying to Z($\nu\bar{\nu}$)V(q$\bar{q}$') in proton-proton collisions at
  $\sqrt{s}$ = 13 TeV}},  \href{http://arxiv.org/abs/2109.08268}{{\tt
  arXiv:2109.08268}}.

\bibitem{CMS:2021fyk}
{\bf CMS} Collaboration, A.~M. Sirunyan et~al., {\it {Search for a heavy vector
  resonance decaying to a ${\mathrm{Z}}_{\mathrm{}}^{\mathrm{}}$ ~boson and a
  Higgs boson in proton-proton collisions at $\sqrt{s} = 13\,\text {Te}\text
  {V} $}},  {\em Eur. Phys. J. C} {\bf 81} (2021), no.~8 688,
  [\href{http://arxiv.org/abs/2102.08198}{{\tt arXiv:2102.08198}}].

\bibitem{Belyaev:2018jse}
A.~Belyaev, A.~Coupe, N.~Evans, D.~Locke, and M.~Scott, {\it {Any Room Left for
  Technicolor? Dilepton Searches at the LHC and Beyond}},  {\em Phys. Rev. D}
  {\bf 99} (2019), no.~9 095006, [\href{http://arxiv.org/abs/1812.09052}{{\tt
  arXiv:1812.09052}}].

\bibitem{Belyaev:2019ybr}
A.~Belyaev, K.~Bitaghsir~Fadafan, N.~Evans, and M.~Gholamzadeh, {\it {Any room
  left for technicolor? Holographic studies of NJL assisted technicolor}},
  {\em Phys. Rev. D} {\bf 101} (2020), no.~8 086013,
  [\href{http://arxiv.org/abs/1910.10928}{{\tt arXiv:1910.10928}}].

\bibitem{ATLAS:2019erb}
{\bf ATLAS} Collaboration, G.~Aad et~al., {\it {Search for high-mass dilepton
  resonances using 139 fb$^{-1}$ of $pp$ collision data collected at
  $\sqrt{s}=$13 TeV with the ATLAS detector}},  {\em Phys. Lett. B} {\bf 796}
  (2019) 68--87, [\href{http://arxiv.org/abs/1903.06248}{{\tt
  arXiv:1903.06248}}].

\bibitem{CMS:2021ctt}
{\bf CMS} Collaboration, A.~M. Sirunyan et~al., {\it {Search for resonant and
  nonresonant new phenomena in high-mass dilepton final states at $ \sqrt{s} $
  = 13 TeV}},  {\em JHEP} {\bf 07} (2021) 208,
  [\href{http://arxiv.org/abs/2103.02708}{{\tt arXiv:2103.02708}}].

\bibitem{ATLAS:2019lsy}
{\bf ATLAS} Collaboration, G.~Aad et~al., {\it {Search for a heavy charged
  boson in events with a charged lepton and missing transverse momentum from
  $pp$ collisions at $\sqrt{s} = 13$ TeV with the ATLAS detector}},  {\em Phys.
  Rev. D} {\bf 100} (2019), no.~5 052013,
  [\href{http://arxiv.org/abs/1906.05609}{{\tt arXiv:1906.05609}}].

\bibitem{CMS:2022yjm}
{\bf CMS} Collaboration, A.~Tumasyan et~al., {\it {Search for new physics in
  the lepton plus missing transverse momentum final state in proton-proton
  collisions at $\sqrt{s}$ = 13 TeV}},
  \href{http://arxiv.org/abs/2202.06075}{{\tt arXiv:2202.06075}}.

\bibitem{ATLAS:2021bjk}
{\bf ATLAS} Collaboration, {\it {Search for high-mass resonances in final
  states with a tau lepton and missing transverse momentum with the ATLAS
  detector}},  ATLAS-CONF-2021-025.

\bibitem{ATLAS:2018tvr}
{\bf ATLAS} Collaboration, {\it {Prospects for searches for heavy $Z^\prime$
  and $W^\prime$ bosons in fermionic final states with the ATLAS experiment at
  the HL-LHC}},  ATL-PHYS-PUB-2018-044.

\bibitem{CMS:2022gho}
{\bf CMS} Collaboration, {\it {Sensitivity projections for a search for new
  phenomena at high dilepton mass for the LHC Run 3 and the HL-LHC}},
  CMS-PAS-FTR-21-005.

\bibitem{Thamm:2015zwa}
A.~Thamm, R.~Torre, and A.~Wulzer, {\it {Future tests of Higgs compositeness:
  direct vs indirect}},  {\em JHEP} {\bf 07} (2015) 100,
  [\href{http://arxiv.org/abs/1502.01701}{{\tt arXiv:1502.01701}}].

\end{thebibliography}\endgroup

\end{document}